\documentclass[pdflatex,sn-nature]{sn-jnl}

\usepackage{graphicx}%
\usepackage{multirow}%
\usepackage{amsmath,amssymb,amsfonts}%
\usepackage{amsthm}%
\usepackage{makecell}
\usepackage{mathrsfs}%
\usepackage[title]{appendix}%
\usepackage[table]{xcolor}%
\usepackage{textcomp}%
\usepackage{manyfoot}%
\usepackage{booktabs}%
\usepackage{algorithm}%
\usepackage{algorithmicx}%
\usepackage{algpseudocode}%
\usepackage{listings}%
\usepackage{caption}
\usepackage{subcaption}
\usepackage{acronym}
\usepackage{orcidlink}
\usepackage{comment}

\newacro{gw}[GW]{gravitational wave}
\newacro{lgwa}[LGWA]{Lunar Gravitational Wave Antenna}
\newacro{lvk}[LVK]{LIGO-Virgo-KAGRA}
\newacro{et}[ET]{Einstein Telescope}
\newacro{snr}[SNR]{signal-to-noise ratio}
\newacro{icrs}[ICRS]{International Celestial Reference System}
\newacro{bbh}[BBH]{binary black hole}
\newacro{nsbh}[NSBH]{neutron star-black hole}
\newacro{bns}[BNS]{binary neutron star}
\newacro{lisa}[LISA]{Laser Interferometer Space Antenna}
\newacro{ssb}[SSB]{Solar System barycenter}
\newacro{cbc}[CBC]{compact binary coalescence}
\newacro{psd}[PSD]{power spectral density}
\newacro{kde}[KDE]{kernel density estimate}
\newacro{ppd}[PPD]{posterior predictive distribution}
\newacro{hpd}[HPD]{highest posterior density}
\newacro{svd}[SVD]{singular value decomposition}

\newcommand{\drm}{{\rm d}}
\newcommand{\irm}{{\rm i}}
\newcommand{\dd}[1]{{\rm d}\, #1}

\newcommand{\LGWAChirpMassWidth}{\(2\times 10^{-4}\ M_{\odot}\)}
\newcommand{\ETChirpMassWidth}{\(4\times 10^{-3}\ M_{\odot}\)}
\newcommand{\LVKChirpMassWidth}{\(8\times 10^{-1}\ M_{\odot}\)}
\newcommand{\LGWAEarlyWarning}{Two minutes} 
\newcommand{\LGWANinetyPercentArea}{65 deg$^{2}$} 
\newcommand{\MinimumUncertaintyFrequency}{0.54Hz}

\newcommand{\defaultwidth}{0.7\columnwidth}

\begin{document}

\title{The geometry of lunar gravitational wave detection}

\author*[1,2,3]{\fnm{Jacopo} \sur{Tissino}~\orcidlink{0000-0003-2483-6710}}\email{jacopo.tissino@gssi.it} 

\author[1,2]{\fnm{Filippo} \sur{Santoliquido}~\orcidlink{0000-0003-3752-1400}}
\author[4]{\fnm{Francesco} \sur{Iacovelli}~\orcidlink{0000-0002-4875-5862}}
\author[5]{\fnm{Ulyana} \sur{Dupletsa}~\orcidlink{0000-0003-2766-247X}}
\author[6]{\fnm{Tito} \sur{Dal Canton}~\orcidlink{0000-0001-5078-9044}}

\author[1,2]{\fnm{Matteo} \sur{Ballelli}~\orcidlink{0000-0003-1512-5423}}
\author[1,2]{\fnm{Ansh} \sur{Chopra}~\orcidlink{0009-0003-5933-4398}}
\author[1,2]{\fnm{Luis Enrique} \sur{Espinosa Castro}~\orcidlink{0009-0004-6030-2788}}
\author[1,2]{\fnm{Laura} \sur{Pezzella}~\orcidlink{0009-0008-6357-6417}}
\author[1,2]{\fnm{Matteo} \sur{Schulz}~\orcidlink{0009-0005-8184-0232}}
\author[1,2]{\fnm{Izumi} \sur{Takimoto Schmiegelow}~\orcidlink{0009-0001-7670-0156}}

\author[1,2]{\fnm{Jan} \sur{Harms}~\orcidlink{0000-0002-7332-9806}}

\affil[1]{\orgname{Gran Sasso Science Institute (GSSI)}, \orgaddress{\city{L'Aquila}, \postcode{I-67100}, \country{Italy}}}
\affil[2]{\orgname{INFN, Laboratori Nazionali del Gran Sasso}, \orgaddress{\city{Assergi}, \postcode{I-67100}, \country{Italy}}}
\affil[3]{\orgname{Theoretisch-Physikalisches Institut, Friedrich-Schiller-Universität Jena}, \orgaddress{Fröbelstieg 1, \postcode{07743} \city{Jena}, Germany}}
\affil[4]{\orgname{William H. Miller III Department of Physics and Astronomy, Johns Hopkins University}, \orgaddress{3400, N. Charles Street,  \postcode{21218} \city{Baltimore}, Maryland, USA}}
\affil[5]{\orgname{Marietta Blau Institute - Austrian Academy of Sciences}, \orgaddress{\postcode{1010} \city{Vienna}, \country{Austria}}}
\affil[6]{\orgname{Université Paris-Saclay, CNRS/IN2P3}, \orgaddress{IJCLab, \postcode{91405} Orsay, \country{France}}}

\abstract{

The \ac{lgwa} is a planned gravitational wave detector on the Moon, targeting the deci-Hertz band and expected to deliver breakthrough discoveries across several science cases, including the Moon's interior structure and astrophysics. In this work, we show that adopting a frame comoving with the \ac{ssb}, but with its origin at a location that minimizes timing uncertainty, reduces the sampling time by an order of magnitude. We present a systematic post-processing procedure to identify the optimal origin within the Solar System for any given signal. We explore alternative timing parametrizations beyond the merger time, and find that they have only a minor impact on parameter uncertainties. Using the stellar-mass black hole binary GW250114 as a case study, we illustrate how these geometrical considerations translate into improved parameter constraints. \LGWAEarlyWarning\  before its merger, the \ac{lgwa} would have measured its chirp mass to a precision of \LGWAChirpMassWidth\ (90\% symmetric) and constrained its sky position to within \LGWANinetyPercentArea\ (90\% \acs{hpd} area); these constraints are tighter than those obtained by the \ac{lvk} detectors, despite a lower signal-to-noise ratio. We connect our results to an analytical approximation proposed by Wen and Chen \cite{wenGeometricalExpressionAngular2010}, which relates the area spanned by the orbital motion of a detector to its efficacy in constraining the sky position of a source. We verify its qualitative validity for compact binary sources with a series of injections, identifying the regimes in which its underlying assumptions break down. Our results demonstrate that inference for long-duration \ac{gw} signals with \ac{lgwa} must be treated as a geometrical problem, in which detector motion, reference-frame choice, and signal evolution jointly determine both parameter constraints and computational efficiency.
}




\maketitle


\section{Introduction}\label{sec:introduction}

The \acf{lgwa} \cite{harmsLunarGravitationalwaveAntenna2021,ajithLunarGravitationalwaveAntenna2024} 
is a gravitational wave detector planned for the next decade. It will employ the response of the Moon to \acp{gw} \cite{weberGravitationalRadiation1967,coughlinConstrainingGravitationalWave2014} with seismic measurements from the lunar pole, where cryogenic operation can occur in the permanently shadowed regions.
While the lunar response is strongest at the normal modes in the milli-Hertz range, thermal noise significantly deteriorates the low-frequency readout: because of this, the best sensitivity for the \ac{lgwa} is achieved in the deci-Hertz range.

The \ac{lgwa} will have significant localization capabilities for sources with a long duration in band, such as neutron star \cite{yelikarBinaryNeutronStars2025}, neutron star-black hole, stellar-mass black hole or white dwarf binaries \cite{benettiObservingDoubleWhite2025}.
This is driven by the measurement of the Doppler shift due to the Moon's motion around the Sun, which can provide precise results even for marginally detected signals.
The long inspiral in band also enables early warning for these compact binaries, ranging from minutes to hours before merger.
This will enable \emph{multi-band} analysis, provided that ground-based detectors such as \ac{et} \cite{punturoEinsteinTelescopeThirdgeneration2010,branchesiScienceEinsteinTelescope2023} are simultaneously operational.

Here we will focus on low-stellar-mass \acp{bbh}, while noting that the science case for the \ac{lgwa} extends much beyond these sources \cite{ajithLunarGravitationalwaveAntenna2024}.
Several authors have previously investigated the low-frequency parameter estimation of stellar-mass binaries in the context of \ac{lisa} \cite{toubianaParameterEstimationStellarmass2020,toubianaDetectabilityParameterEstimation2022,marsatExploringBayesianParameter2021,wuMultibandParameterEstimation2025}. 

Some of the authors \cite{iacovelliGravitationalwaveParameterEstimation2025} have performed parameter estimation with the \ac{lgwa} for a heavy binary, chosen to be compatible with the heaviest observed so far, GW231123 \cite{abacGW231123BinaryBlack2025}. There, we noted how its long observation can yield tight constraints on certain parameters, despite a comparatively low \ac{snr}. However, the role of geometry in the inference process, particularly the impact of reference-frame choices and timing parametrizations, has not yet been systematically explored.

In this paper, we address this question using a recently observed black hole coalescence as an example: 
GW250114 \cite{abacGW250114TestingHawkings2025}, the clearest gravitational wave signal detected to date.
Its intrinsic parameters and distance were similar to those of the first detection, GW150914 \cite{ligoscientificcollaborationandvirgocollaborationObservationGravitationalWaves2016}. 

Sources similar to GW250114, \emph{i.e.}, binaries of slowly-spinning black holes with component masses around $35M_{\odot}$, are relatively common \cite{callisterObservedGravitationalWavePopulations2024}: at design sensitivity, we expect the \ac{lgwa} to detect  at least some. This makes signals like GW250114 a suitable benchmark for LGWA observations..

While the signal lasted less than a second for the \ac{lvk} detectors \cite{abacGW250114TestingHawkings2025}, it would have spent several months inspiraling in the frequency band of \ac{lgwa}.
This gives rise to precise estimates 
of its sky location and mass, 
despite the lower \ac{snr}.

A central result of this work is that the choice of a reference position used to parametrize the signal arrival time becomes an essential part of the inference process,
which differentiates this type of analysis 
from both ground-based and \ac{lisa}-like detectors.
For ground-based detector networks, the timing uncertainty is minimized by a point within the Earth, typically near the detector with the highest \ac{snr} \cite{rouletRemovingDegeneracyMultimodality2022}, and this point can be approximated as an \ac{snr}-weighted average of the detector locations \cite{singerRapidBayesianPosition2016}.
The standard choice of a frame with origin at the \ac{ssb} leads to inefficient sampling. 
We show that, by considering a frame comoving with the \ac{ssb} one can identify an optimal reference position (equivalently, the reference frame's origin) that minimizes timing uncertainty, while ensuring a proper description of the orbital Doppler modulation. This optimal point can be determined systematically in post-processing.

A related question to ask is which timing parameter is most naturally constrained: the coalescence time is not a natural metric, as it is out of the detector band. 
We explore the effect of measuring time at a reference frequency within the deci-Hertz band, which results in a lower variance in the timing parameter. 

We also give a qualitative understanding of the way the \ac{lgwa} localizes, showing that it is mainly driven by the measurement of the Doppler phasing during the Moon's orbit around the Sun.
For a monochromatic source at a frequency \(f_0\), observed for a duration \(T\) with a total \ac{snr} of \(\rho_T\), the following Fisher-matrix expression for the angular resolution can be computed \cite{wenGeometricalExpressionAngular2010}: 
\begin{equation} \label{eq:wen-chen-area}
    \Delta \Omega \approx \frac{c^{2}}{(f_0 \rho_{T})^{2} A_s(T)}\,.
\end{equation}

The quantity \(A_s(T)\) is the area spanned by the detector's orbit in a period \(T\), weighted by the SNR accumulation during the motion.
This area determines the angular resolution; for GW250114, most of the \ac{snr} is accumulated in a short fraction of the total observation, in its last days, meaning that \(A_s\) is significantly smaller than the geometrical area of the orbit.
%

This paper is organized as follows. 
In section \ref{sec:methods}, we describe the waveform model, the LGWA response, and the likelihood used in our analysis. 
In section \ref{sec:injections}, we present the results of injection studies based on a GW250114-like signal. 
We first discuss parameter constraints and their comparison with ground-based detectors (section \ref{sec:constraints}). 
We then introduce a post-processing framework to explore different reference positions and timing parametrizations (section \ref{sec:shifting-timing}), and develop an analytic understanding of the resulting timing uncertainties (section \ref{sec:analytic-timing}). 
In section \ref{sec:sampling-time}, we quantify the impact of these choices on sampling efficiency. 
In section \ref{sec:localization-function-area}, we investigate the geometrical scaling of sky localization with the detector's orbital area, comparing it to the analytical results in Ref.~\cite{wenGeometricalExpressionAngular2010}, which we rederive and expand upon in appendix \ref{app:angular-resolution-derivation}.
Finally, in section \ref{sec:conclusions}, we summarize our findings and discuss their implications for future lunar GW observations.

\section{Methods}\label{sec:methods}

\subsection{Waveform modelling and priors}\label{sec:waveforms}

We parameterize our waveforms through a set of intrinsic parameters \(\theta _{\text{int}} = \left\lbrace \mathcal{M}, q, \chi_1, \chi_2 \right\rbrace\), as well as extrinsic parameters \(\theta _{\text{ext}} = \left\lbrace  d_L, \mathtt{ra}, \mathtt{dec}, \phi_0, \iota, \psi, t_{\text{origin}} \right\rbrace\). This section and the next will discuss the meaning of these parameters, how they affect the detected signal, and the priors used for them during sampling.

In the frequency domain, the two polarizations \(h_+\) and \(h_\times\) which compose the gravitational waveform emitted by a binary system can be decomposed into spin-weighted spherical harmonics as follows:
\begin{align}\label{eq:polarizations-all-modes}
h_+ (f) - i h_\times (f) = \frac{1}{d_L} \sum _{l=2}^{\infty } \sum
_{m=-\ell}^{\ell} h_{\ell m} (f; \theta_{\text{int}})\, {}^{(-2)}Y_{\ell m} (\iota, \phi_0)
\,.
\end{align}

The extrinsic parameters that appear 
in Equation~\ref{eq:polarizations-all-modes} 
are the luminosity distance \(d_L\), 
the reference phase \(\phi_0\) 
and the inclination angle \(\iota\).

At low frequencies and for nearly-equal masses, 
the \(\ell = |m| = 2\) mode dominates this sum, 
while the other ones only gain relevance close to merger, 
see \emph{e.g.}\ ref.\ \cite[fig.\ 1]{marsatExploringBayesianParameter2021}.
The upper bound on frequency in our analysis is 3 Hz, 
while the merger and ringdown of GW250114 occurred
at frequencies exceeding 100 Hz.
For this reason, as well as for simplicity, 
we only consider the dominant mode throughout this work.
In this case, the expressions for the two polarizations simplify to 
\begin{align}
  h_+(f) &= \frac{1 + \cos^2 \iota}{2 D_L} \sqrt{ \frac{5}{4 \pi}} h_{22}(f, \theta_{\text{int}}) \\
  h_\times(f) &= \frac{\cos \iota}{D_L} \sqrt{ \frac{5}{4 \pi}} h_{22}(f, \theta_{\text{int}}) e^{i \pi / 2} 
\,.
\end{align}

The intrinsic waveform, \(h_{22}(f, \theta_{\text{int}})\), depends on 
the masses of the sources \(m_1\) and \(m_2\) (with \(m_1>m_2\)), 
which we express in terms of the chirp mass \(\mathcal{M} = (m_1 m_2)^{3/5} / (m_1 + m_2)^{1/5}\) and the mass ratio \(q = m_2 / m_1\).
Throughout this work we refer to detector-frame masses, 
which are related to the source-frame ones, 
denoted as \(m_\text{source}\), by \(m = (1+z) m_\text{source}\), where \(z\) is the cosmological redshift. 

We adopt a uniform prior in the component masses, in the region defined by \(\mathcal{M}_0 - 6 \times 10^{-3} M_\odot \leq \mathcal{M} \leq  \mathcal{M}_0 + 6 \times 10^{-3} M_\odot\) and \(0.125 \leq q \leq 1\), also constraining the individual masses to \(5 M_\odot \leq m_i \leq 100 M_\odot\) for \(i = 1, 2\).
Here, \(\mathcal{M}_0\) denotes the injected value of the chirp mass.
The narrow prior on this parameter is chosen for computational efficiency, since the large number of cycles in band gives rise to an extremely precise posterior.

The precessing spin parameter \(\chi_p\) for GW250114 was constrained to be \(\lesssim 0.2\) (or \(0.3\), depending on the waveform model) at 90\% credibility \cite{abacGW250114TestingHawkings2025}.
The impact of spin precession for this event in the deci-Hertz band could have been measurable, but for simplicity we also assume the spins of the two black holes, \(\vec{\chi}_1\) and \(\vec{\chi}_2\), were aligned to the orbital angular momentum.
Our parameters are then only the dimensionless aligned spin components \(\chi_{1z}\) and \(\chi_{2z}\). 
This also implies that the orbital and total angular momentum of the binary are aligned, therefore the inclination angle \(\iota\) is constant and equal to the commonly used parameter \(\theta_{\text{JN}}\), the angle between the observation direction and the total angular momentum.
For each of these spin components, we adopt a derived prior, obtained by projecting a prior on the spin vector, isotropic and uniform in its magnitude, onto the \(z\) axis. 
This makes our results on this spin component comparable to those of the \ac{lvk}, and it is equivalent to the following expression in each spin magnitude \cite[eq. A7]{langeRapidAccurateParameter2018}: \(p(\chi_{i, z}) = - \ln ( |\chi_{i, z} / \chi_z^{\text{max}}|) / 2 \chi_z^{\text{max}}\), and we take \(\chi_z^{\text{max}} = 0.9\).

In principle, GW250114 also could have exhibited significant eccentricity in the \ac{lgwa} band: the eccentricity parameter was constrained to \(e \leq 0.03\) at 13.33 Hz under a uniform prior \cite{abacGW250114TestingHawkings2025}, but since gravitational wave emission dissipates eccentricity over time this does not rule out detectable eccentricities at lower frequencies.
Measuring eccentricity at low frequency is indeed 
a significant advantage of a mission such as the \ac{lgwa}.
Nevertheless, we leave this investigation for future work, and assume that the binary was quasicircular at all times.

We approximate the intrinsic waveform with a frequency domain model from the \texttt{TaylorF2} family \cite{buonannoComparisonPostNewtonianTemplates2009,messinaQuasi55PNTaylorF2Approximant2019}, including spin-spin and spin-orbit terms accurate to 3.5 post-Newtonian order in amplitude and phase.
We re-implement this model and accelerate it with \texttt{numba} \cite{lamNumbaLLVMbasedPython2015}.

Our prior is uniform on the sphere for right ascension and declination, a sine distribution for the inclination angle \(\iota = \theta_{JN}\) (such that \(\cos \iota \sim \mathcal{U}(-1, 1)\)), and uniform in comoving volume and source frame time for the luminosity distance \(d_L\); we adopt values for the cosmological parameters obtained from Planck 2018 cosmological results \cite{Planck:2018vyg}.
We adopt uniform prior distributions for all remaining parameters: \(\psi \sim \mathcal{U}(0, \pi)\), \(\phi_c \sim \mathcal{U}(0, 2 \pi)\),  and \(t_{\text{origin}} \sim \mathcal{U}(t_0 - 2\text{s}, t_0 + 2\text{s})\), where \(t_0\) represents the injected value.
Section \ref{sec:response} will give more details on how the parameter \(t_{\text{origin}}\) is precisely defined.

\subsection{The LGWA response}\label{sec:response}

When a gravitational wave reaches the Moon, the lunar surface and interior react elastically,
resulting in differential displacement between them and an inertial observer \cite{weberDetectionGenerationGravitational1960,weberGravitationalRadiation1967,dysonSeismicResponseEarth1969,coughlinConstrainingGravitationalWave2014,belgacemCouplingElasticMedia2024,biResponseMoonGravitational2024}.
This is the detection principle for the \ac{lgwa} \cite{harmsLunarGravitationalwaveAntenna2021}, which will measure horizontal displacement along two horizontal directions we refer to as \emph{channels}.

For the following discussion, we will work in a fixed reference frame with axes aligned to the \ac{icrs} frame.
The origin of this reference frame may be any point in the Solar System; we do however require it to be stationary with respect to the SSB, so the position of the \ac{lgwa}, the center of the Moon or of the Earth are not valid candidates. Section \ref{sec:shifting-timing} will describe how this choice can be explored based on the results from a run, while table \ref{tab:timing-uncertainty} anticipates the effect this choice can have on the uncertainty on the timing parameter.

A gravitational wave coming from a single direction in the sky can be written as \cite[section 7.2]{maggioreGravitationalWavesVolume2007}
\begin{align}
h_{ij} = h_+ (t) e_{ij}^{+} + h_\times (t) e_{ij}^{\times}
\,,
\end{align}
where the two basis tensors \(e_{ij}^{+}\) and \(e_{ij}^{\times}\) are written as 
\begin{align}
e_{ij}^{+} = u_{i}u_{j} - v_{i}v_{j}
\qquad \text{and}
\qquad
e_{ij}^{\times} = u_{i}v_{j} + v_{i}u_{j}\,.
\end{align}

The unit vectors \(u\) and \(v\), together with unit vector \(m\) which defines the sky position of the source,
form an orthonormal reference frame, defined by three angles: a longitude \(\phi\) and a colatitude \(\theta\) (right ascension and the complementary to the declination), as well as a polarization angle \(\psi \in [0, \pi]\) which parametrizes rotations around the \(m\) axis.
These vectors are always constant for a given source, as motion within the Solar System is negligible compared to the source distance.

Explicitly, we write these vectors as:
\begin{align}
m &= \begin{pmatrix}
\sin \theta \cos \phi  \\
\sin \theta \sin \phi \\
\cos \theta
\end{pmatrix} \label{eq:propagation-unit-vector}\\ 
u &= \begin{pmatrix}
-\sin \phi \\
\cos \phi \\
0
\end{pmatrix} \cos \psi
- \begin{pmatrix}
\cos\theta \cos \phi \\
\cos\theta \sin \phi \\
-\sin\theta
\end{pmatrix}\sin \psi \\
v &= 
\begin{pmatrix}
-\sin \phi \\
\cos \phi \\
0
\end{pmatrix} \sin \psi
+ \begin{pmatrix}
\cos\theta \cos \phi \\
\cos\theta \sin \phi \\
-\sin\theta
\end{pmatrix}\cos \psi \,.
\end{align}

We now turn to modelling the response of a single seismometer channel to the gravitational wave \(h_{ij}(t)\), that is, the differential motion along a horizontal axis \(b\) along the lunar surface.

If the gravitational wave signal \(h_{ij}(t)\) has a frequency \(f\), this can be approximately written as the contraction of the gravitational strain tensor with an antenna pattern tensor \(\hat{D}_ij\) multiplied by a frequency-dependent response function \(L(f)\), 
\begin{equation} \label{eq:response-tensorial}
    s(t_\text{det}, f) = h_{ij}(t) \hat{D}_{ij}(t_\text{det}) L(f)\,.
\end{equation}

We express the strain $h_{ij}(t)$ as a time series, shifted in time as needed so that some evolutionary stage of the signal happens when $t=0$.
This stage is typically chosen to be the merger in ground-based \ac{cbc} analyses: this is not the only option; and we will later show that it is in fact suboptimal for the analysis considered in this work.

Since the \(\ell |m| = 22\) mode of a quasicircular signal rises monotonically in frequency, we can define a one-to-one time-frequency mapping \(t_{22}(f; \theta_{\text{int}})\) \cite{marsatExploringBayesianParameter2021}.  
The time \(t_{22}(f; \theta_{\text{int}})\) for some fixed frequency \(f\) can then be used as the reference evolutionary stage. 

It is then straightforward to relate this parameterization to the standard one at the merger: in the stationary phase approximation, the time-frequency mapping can be made explicit as
\begin{align} \label{eq:time-to-merger}
  t_{\text{merger}} - t_{22}(f;\theta_{\text{int}}) = \frac{1}{2 \pi} \frac{ \text{d} \varphi_{22}(f; \theta_{\text{int}})}{ \text{d} f} \approx - \frac{5}{256 \pi^{8/3}} \mathcal{M}^{-5/3} f^{-8/3}
\,,
\end{align}
where \(\varphi_{22}\) is the phase of the \(\ell |m| = 22\) mode. 
The expression in the last approximate equality only considers the lowest order of emission, but for all practical calculations we use the same 3.5PN approximation as in the computation of the waveform \cite{buonannoComparisonPostNewtonianTemplates2009,messinaQuasi55PNTaylorF2Approximant2019}.

This choice of evolutionary stage is not unique nor general. 
In the case of an eccentric binary, for example, the time-to-frequency mapping is not monotonic, so a generalization is required.
This non-uniqueness is not a conceptual problem: even for a generic signal, including eccentricity and higher order modes beyond the \(\ell |m| = 22\) one, we just need to choose a monotonic function of the orbital dynamics, which could for example be the orbit-averaged frequency of the 22 mode. 
Some choices might provide more stable results than others, 
we shall investigate this in future work.

Let us denote with $t_{0}$ the time at which this stage reaches the origin of the coordinate frame. 
Then, we will have the following relation for the time at the detector \(t _{\text{det}}\):
\begin{equation}
t_\text{det} = t + t_{0} - \frac{\vec{m}\cdot \vec{r}(t_\text{det})}{c}\,,
\end{equation}
where $\vec{r}$ is the vector denoting the position of the detector as measured from the origin of the reference frame.

This expression is valid under the assumption that, for a stationary observer located at the origin of the reference frame, the waveform is unaffected by Doppler shifts. 
The motion of the Moon around the Sun during a year of observation is not negligible: the reference frame used for our analyses must be comoving with the \ac{ssb}.
This does not necessarily mean the \emph{origin} should be the \ac{ssb} itself: it can be any fixed point in the Solar System. 
Indeed, as we shall discuss in section \ref{sec:shifting-timing}, choices other than the \ac{ssb} are preferrable for the analysis of stellar-mass black holes like GW250114 with the \ac{lgwa}.

We write the detection tensor for each of the two LGWA channels in the form
\begin{equation} 
\hat{D}_{ij}(t_{\text{det}}) = n_{i}(t_{\text{det}}) b_{j}(t_{\text{det}})\,,
\end{equation}
where $n$ is the normal to the lunar surface at the detector location at time $t$, while $b$ is a vector parallel to the lunar surface and aligned to the measurement channel.
These are time-dependent, since the Moon will in general not be stationary for the duration of the observation.
This expression is approximate and derived based on the Dyson model of lunar response \cite{dysonSeismicResponseEarth1969}, in which gravitational wave strain gives rise to an effective force term dependent on the gradient of the shear modulus \(\mu\), as \(f_i = h_{ij} \nabla_j \mu(\vec{r})\) \cite{harmsLunarGravitationalwaveAntenna2021,biResponseMoonGravitational2024}.
The dominant contribution to the shear modulus' gradient near the surface is expected to be the surface itself, hence the normal unit vector.

Finally, the response function \(L(f)\) describes the displacement corresponding to a unit monochromatic strain injected at a frequency \(f\). 
Its value in the deci-Hertz band is expected to be on the order of \(10^{5}\)m \cite{biResponseMoonGravitational2024}, though this is still a topic of discussion in the literature, with full 3D simulations only recently becoming available \cite{zhangNumericalSimulationLunar2025,zhangThickLunarCrust2026}. 
For the purposes of this work, a fiducial model was assumed, as described in section \ref{sec:likelihood}.

In practice, the tensor products in Equation~\ref{eq:response-tensorial} simplify to scalar products between vectors as follows:
\begin{align}
\frac{s(t, f)}{L(f)} = h_{+}(t) \underbrace{((n\cdot u)(b\cdot u)-(n\cdot v)(b\cdot v))}_{F_{+}} 
+
h_{\times}(t) \underbrace{((n\cdot u)(b\cdot v)+(n\cdot v)(b\cdot u))}_{F_{\times}}
\,.
\end{align}

We accelerate the computation of the response by caching the position of the Moon, \(r(t _{\text{det}})\), as well as angles defining the orientation vectors \(n(t _{\text{det}})\) and \(b(t _{\text{det}})\), on a fixed grid.
These vectors are computed thanks to the \texttt{lunarsky} package \cite{ACTON20189,annexSpiceyPyPythonicWrapper2020,8477061,ACTON199665}.

An independent implementation of this response was developed within the \texttt{gwfast} package \cite{iacovelliForecastingDetectionCapabilities2022} for the analysis of the event GW231123. 
We have verified that results obtained with the two implementations agree with each other. 

\subsection{The LGWA likelihood}\label{sec:likelihood}

\begin{figure}[ht]
\centering
\includegraphics[width=\columnwidth]{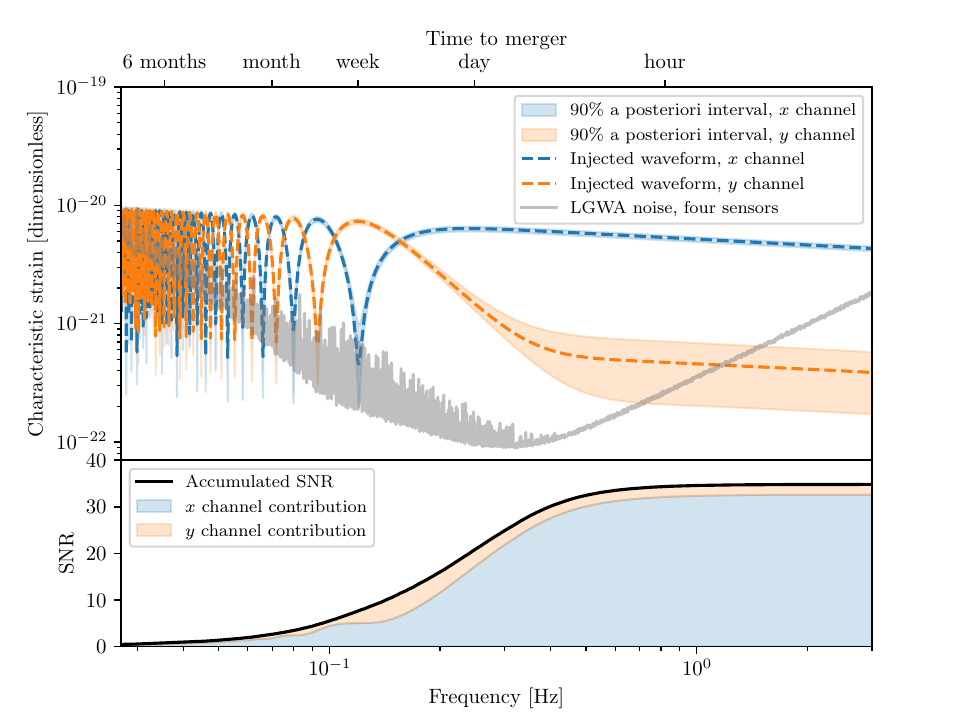}
\caption{Top panel: Projected characteristic strain amplitude for the GW250114-like injection, shown together with the 90\% interval from the recovered \ac{ppd}. Black vertical lines denote the frequency band which was used for the analysis: 0.027 Hz to 3 Hz, corresponding to one year of observation. The modulations are due to the rotation of the Moon on its axis.
Bottom panel: \ac{snr} accumulation for the same injection, showing the contributions from both \ac{lgwa} channels.}
\label{fig:ppd-full-bandwidth}
\end{figure}

We assume that the displacement noise in our seismometers can be described as a stationary, colored Gaussian process:
then, its distribution can be described in the Fourier domain through Whittle likelihood. 
In terms of the displacement data \(s(f)\) and the predicted \ac{gw} displacement \(h_s(f; \theta)\), this reads:
\begin{align} \label{eq:likelihood-displacement-space}
\log \mathcal{L}(s | \theta) = - \frac{1}{2} \int \frac{|s(f) - h_s(f; \theta)|^2  }{S_s(f)} \,\text{d}f + \text{const.}
\,,
\end{align}
where \(S_s(f)\) is the power spectral density of the noise in the displacement measurement.

When working with real data, the displacement \(s\) would arise from the true lunar response \(L^{\text{true}}(f)\), while the predicted displacement \(h_s(f; \theta)\) would be proportional to the response predicted by our response model \(L^{\text{model}}(f; \lambda)\) depending on some parameters \(\lambda\).

At the moment, however, we perform our injection studies by fixing the lunar response to a fiducial value \(L^{\text{true}}(f) = L^{\text{model}}(f) = \hat{L}(f)\), and assuming it is known during inference. 
If this is the case, this response can be factored out of the numerator in~\ref{eq:likelihood-displacement-space} and absorbed in the power spectral density, to yield the strain space density \(S_h(f) = S_s(f) / \hat{L}^2(f)\).
 This is what has been typically shown in the literature \cite{ajithLunarGravitationalwaveAntenna2024,vanheijningenPayloadLunarGravitationalwave2023}.

We perform parameter estimation exploring the distribution of the following log-likelihood ratio:
\begin{align} \label{eq:likelihood-strain-space}
\log \mathcal{L}(d_h | \theta) &= \Re \langle d_h | h(\theta)\rangle - \frac{1}{2} \langle h(\theta) | h(\theta)\rangle\,,
\end{align}
written in terms of the noise-weighted scalar product (Equation~\ref{eq:innProd}).

This quantity is the logarithm of the ratio between the likelihoods in the signal versus the noise hypothesis. 
The minimum and maximum frequency fixed to $27\text{ mHz}$ and 3 Hz respectively; they are chosen so that the fiducial signal's duration is one year.
As Figure \ref{fig:ppd-full-bandwidth} shows, they roughly correspond to when the signal lies in the sensitivity ``bucket'' of the \ac{lgwa}. 
These limits are conservative: decreasing the minimum integration frequency would lead to improved constraints.
However, that would correspond to searching a significantly longer duration of \ac{lgwa} data for an extremely weak signal.

We compute this likelihood with the relative binning approach \cite{zackayRelativeBinningFast2018}, as well as analytic phase marginalization \cite{LIGOT1300326v1AnalyticMarginalisation,thraneIntroductionBayesianInference2019}. Details on the frequency grid and convergence validation are discussed in appendix \ref{app:relative-binning}.

\section{Injection results}\label{sec:injections}

We perform a zero-noise injection corresponding to the maximum likelihood point for GW250114 for the \ac{lgwa}, with the parameter values reported in table \ref{tab:injection-parameters}.
We use the parameters, priors, and likelihood described in section \ref{sec:methods},
and only simulate the signal for its last year from merger, which encompasses most of the \ac{snr} for this source as shown in Figure \ref{fig:ppd-full-bandwidth}.
While this point was obtained with a different waveform approximant than the one we are using, namely \texttt{NRSur7dq4} \cite{varmaSurrogateModelsPrecessing2019}, we expect comparable results at such low frequencies.

Due to the long duration of the signal, the detector's motion is not negligible, and the choice of timing parameter can significantly affect sampling efficiency.
In later sections of this work (\ref{sec:shifting-timing}) we will discuss this in detail; here we limit ourselves to describing the heuristic parameter choice we adopted for all preliminary analysis.
We use the merger time as our parameter, and we set the origin of the reference frame on the position of the \ac{lgwa} when it receives the merger. 
We compute this exactly based on the injected signal.

Since the timeline for the construction of the \ac{lgwa} is very roughly comparable to that of the \ac{et}, 
we also perform two injections with the same parameters
and its two proposed configurations: a triangle and two L-shaped detectors, both in Europe \cite{branchesiScienceEinsteinTelescope2023}.

We give more details about the injections with the \ac{lgwa} in appendix \ref{app:injection-details}, and those with \ac{et} in appendix \ref{app:et-injections}.

\subsection{Parameter constraints} \label{sec:constraints}

The early inspiral measured by the \ac{lgwa} consists of a very large number of cycles, 
and it is significantly affected by 
the modulation driven by the detector's motion, predominantly constituted by the lunar orbit around the Sun.
This gives rise to tight constraints 
on the parameters which determine the phase: chirp mass and sky position \cite{iacovelliGravitationalwaveParameterEstimation2025}, despite an overall lower \ac{snr} for this signal: around 35, as opposed to 
the value of 77--80 reported by the \ac{lvk} \cite{abacGW250114TestingHawkings2025}.

\begin{figure}[ht]
    \centering
    \includegraphics[width=\columnwidth]{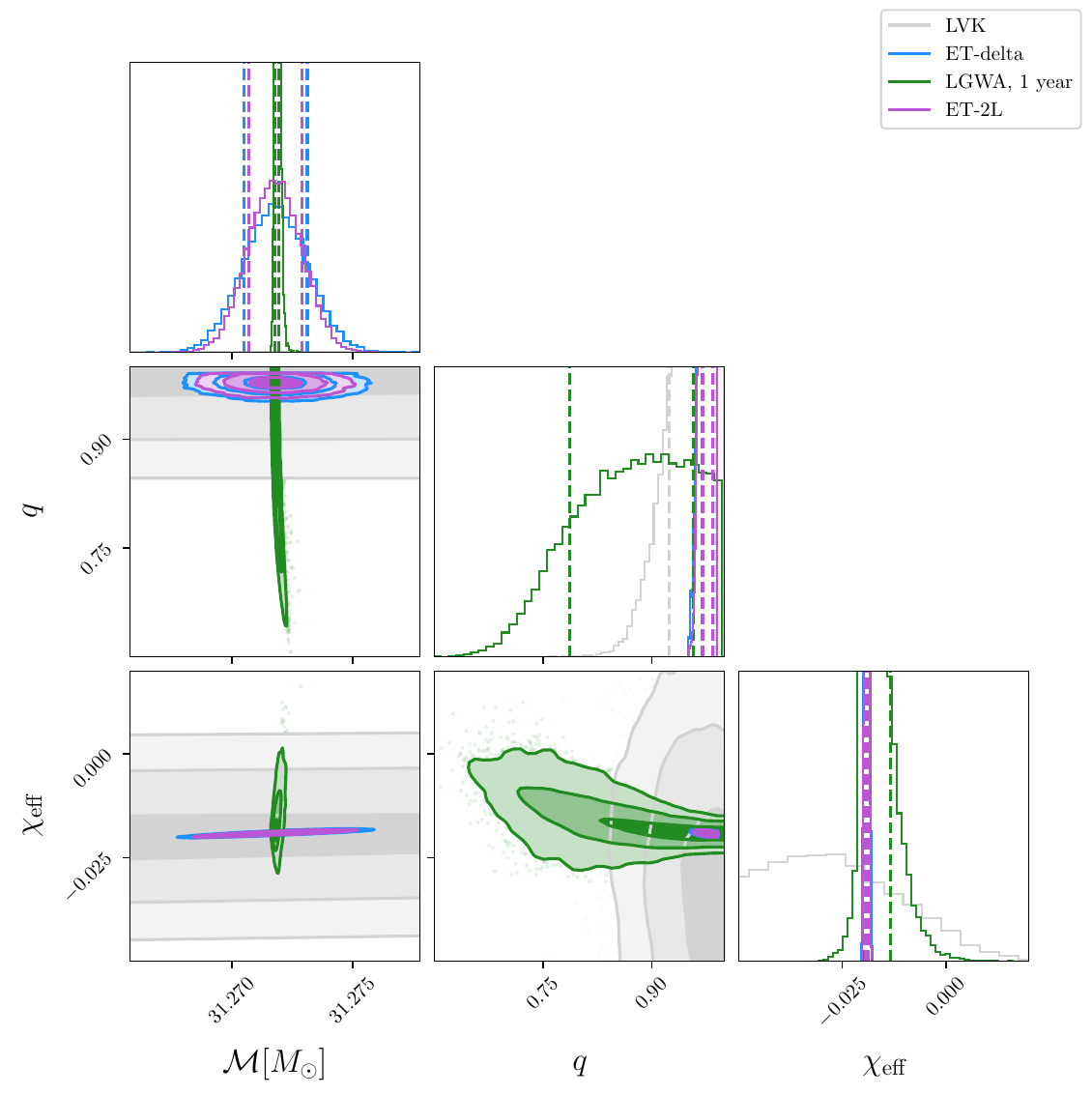}
    \caption{Posterior distributions on chirp mass, mass ratio and effective spin parameter for the injections described in section \ref{sec:injections}with the \ac{lgwa} and \ac{et} in two configurations, as well as the posterior obtained by the \ac{lvk} using real data.}
    \label{fig:mchirp-q-constraint}
\end{figure}

\begin{figure}[ht]
    \centering
    \includegraphics[width=\columnwidth]{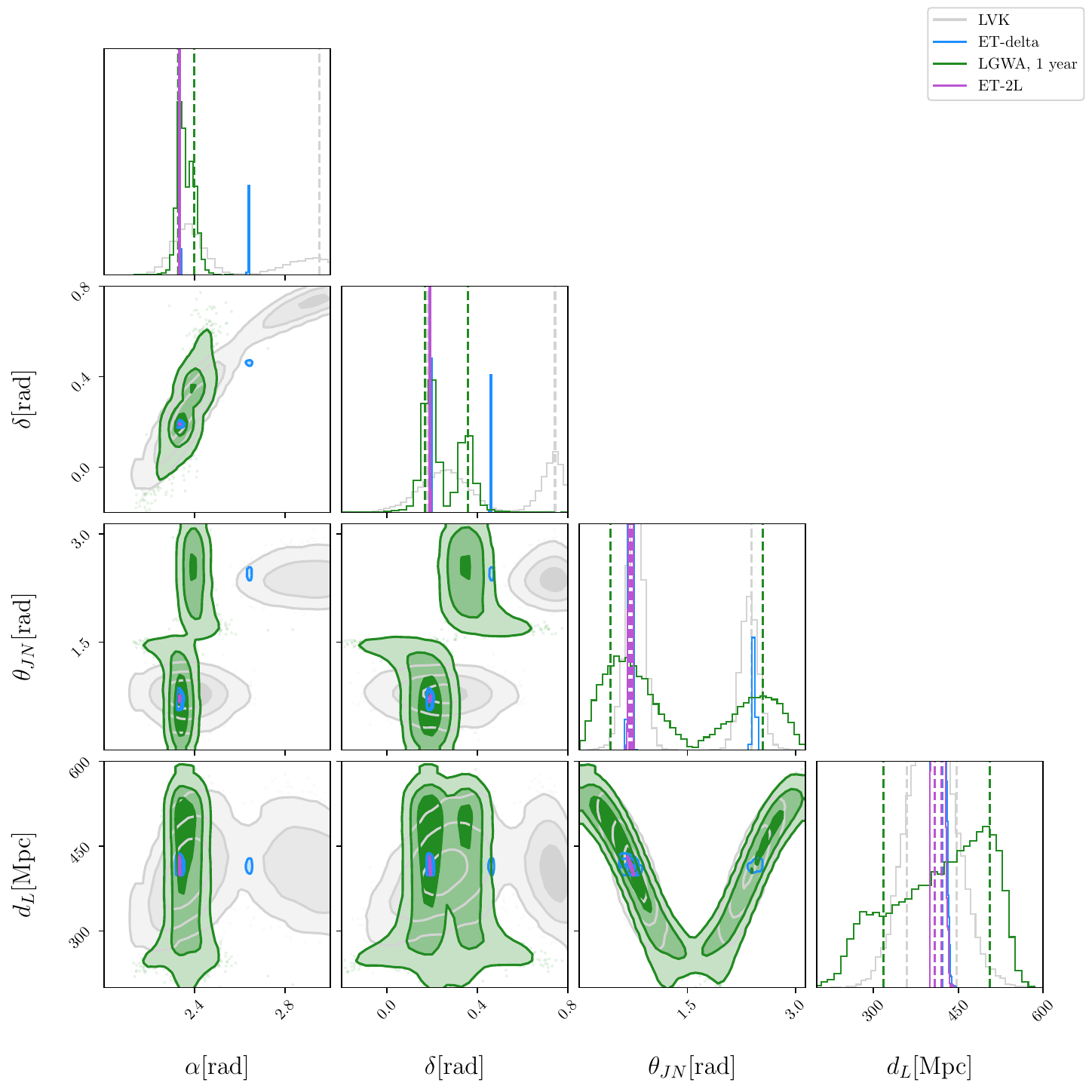}
    \caption{Posterior distributions on sky position, inclination angle and distance for the injections described in section \ref{sec:injections} with the \ac{lgwa} and \ac{et} in two configurations, as well as the posterior obtained by the \ac{lvk} using real data.}
    \label{fig:sky-loc-constraint}
\end{figure}

In Figure \ref{fig:mchirp-q-constraint} we show a comparison of the posterior distributions on chirp mass and mass ratio for these configurations.
The posterior obtained with the \ac{lgwa} is extremely tight, with a width of \LGWAChirpMassWidth\ of the 90\% credible interval. 
The Einstein Telescope performs similarly in either of its configurations, with a width of \ETChirpMassWidth, while the analysis of the real event's data by the \ac{lvk} collaboration constrained it to a width of \LVKChirpMassWidth---a precise measurement in its own right.
While the chirp mass affects the signal at all frequencies, and is thus well-constrained by observations with many cycles in any band,
the mass ratio predominantly affects the high-frequency portion of the waveform, therefore the constraint we obtain on it with the \ac{lgwa} is weaker than that of ground-based detectors.
As the two posteriors extend in different directions, they provide complementary information: combining lunar and terrestrial detectors, we can obtain exquisite precision on the component masses of the binary.

The spins, like the mass ratio, affect the waveform at high frequency; however in this case the early inspiral observed by the \ac{lgwa} provides a tighter constraint than that of the \ac{lvk}.

The sky location posterior for this injection is bimodal, as shown in Figure \ref{fig:sky-loc-constraint}, with a 90\% area of \LGWANinetyPercentArea.
The two modes are close to each other in the sky, and they correlate with face-on and face-off values for the inclination angle; the same phenomenon is observed in the \ac{lvk} posterior.

For comparison, the sky area for the GW250114 event as detected by the two LIGO detectors was 145 square degrees \cite{ligoscientificcollaborationLIGOVirgoKAGRA2025}, in the shape of a thin ring as is typical with two interferometers \cite{fairhurstTriangulationGravitationalWave2009}.

\subsection{Shifting the timing parameterization in post-processing} \label{sec:shifting-timing}

The injections we performed used a reference frame with origin at a position \(r_0\), and used the time at merger \(t_0\) as a reference.

Once we have posterior samples from an analysis performed in this manner, we can post-process them to obtain samples which refer to a new position \(r\), as 
\begin{align}
t_{i}^{\text{mrg}, r} = t_{i}^{\text{mrg}, r_0} - \frac{(r-r_0) \cdot n_{i}}{c}
\,.
\end{align}

We can also shift our parameterization from the time at merger to the time when the ($\ell |m| = 22$ mode of the) signal reaches an arbitrary frequency \(f\).
This is given by Equation \ref{eq:time-to-merger}.

A low-frequency observation can constrain \(t(f)\) better than \(t^{\text{mrg}}\), for some values of \(f\) in the band of the detector.

Combining these contributions, the \(i\)-th post-processed timing posterior sample at a position \(r\) and a frequency \(f\) is given by
\begin{align} \label{eq:timing-samples-shift}
t_{i}^{f, r} = t_{i}^{\text{mrg}, r_0} 
- \frac{(r-r_0 ) \cdot \hat{n}_{i}}{c}
+ \frac{1}{2 \pi} \frac{ \text{d} \varphi_{22}(f; \theta_i)}{ \text{d} f}
\,.
\end{align}

The uncertainty of these alternative parameterizations for time can be significantly lower than the one for \(t^{\text{mrg}}\), as shown in Table \ref{tab:timing-uncertainty} for some examples.

\begin{table}
\begin{tabular}{ccc}
\hline
Location & Stage & Timing uncertainty (90\% interval) \\
\hline
\rowcolor{green!15} global optimum (see fig \ref{fig:timing-by-location}) & \(f_{22}=0.56\) Hz & 0.1s \\
\rowcolor{red!4} Earth & \(f_{22}=0.56\) Hz & 0.15s \\
\rowcolor{red!6} Earth & merger & 0.24s \\
\rowcolor{red!6} Moon & merger & 0.25s \\
\rowcolor{red!31} Solar System Barycenter & merger & 11.66s \\
\rowcolor{red!31} Solar System Barycenter & \(f_{22}=0.56\) Hz & 11.72s \\
\rowcolor{red!50} Earth & \(f_{22}=0.03\) Hz & 238s \\
\hline
\end{tabular}
\caption{Uncertainty in timing based on the position
of the origin of the reference frame and the reference evolutionary stage. The Earth and Moon are not comoving with the \ac{ssb}, and are therefore not valid origins: when used as labels, they are shorthand for their locations when they receive the stage in the given row.}
\label{tab:timing-uncertainty}
\end{table}

\begin{figure}[ht]
\centering
\includegraphics[width=\columnwidth]{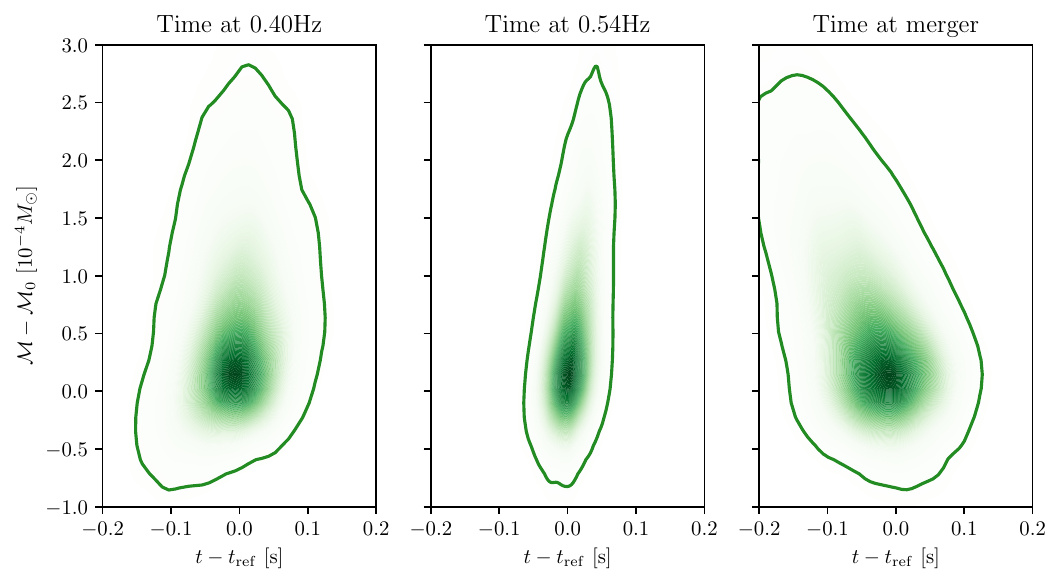}
\caption{Posterior distributions on chirp mass and timing parameter, for different choices of reference evolutionary stage. The shading is a \ac{kde} recontruction of the distribution, while the contour is 90\% credible area.
The asymmetry in the chirp mass distribution is due to railing against the \(q=1\) constraint. 
In all panels, time is measured at the same location in the Solar System, the global optimum, shown in the central row of Figure \ref{fig:timing-by-location}.
}
\label{fig:mchirp-time}
\end{figure}

\begin{figure}[ht]
\centering
\includegraphics[width=\columnwidth]{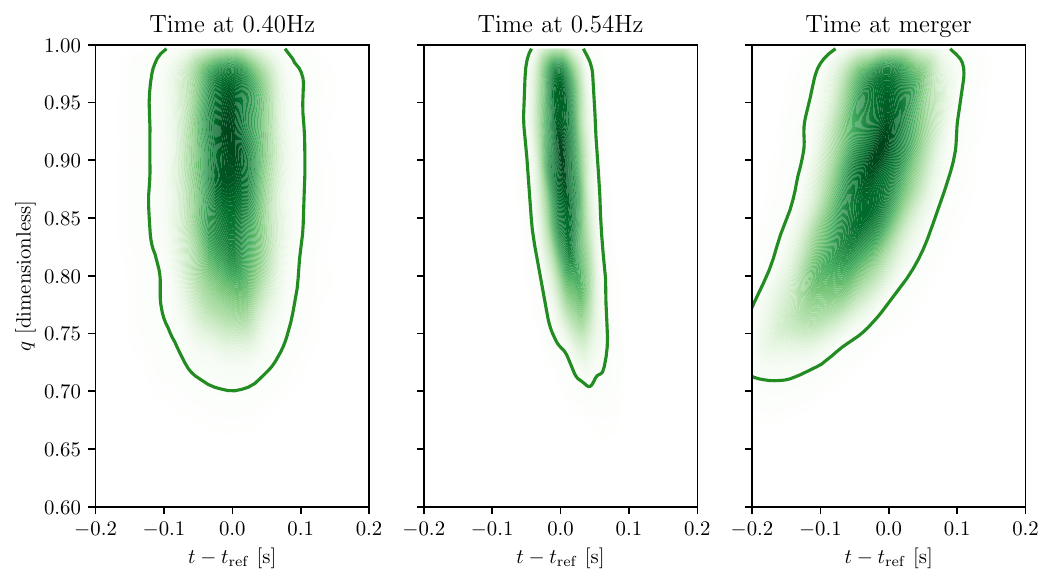}
\caption{Posterior distributions on mass ratio and timing parameter, for different choices of reference evolutionary stage, with the same scenarios depicted in Figure \ref{fig:mchirp-time}.}
\label{fig:q-time}
\end{figure}

\begin{figure}[ht]
\centering
\includegraphics[width=\columnwidth]{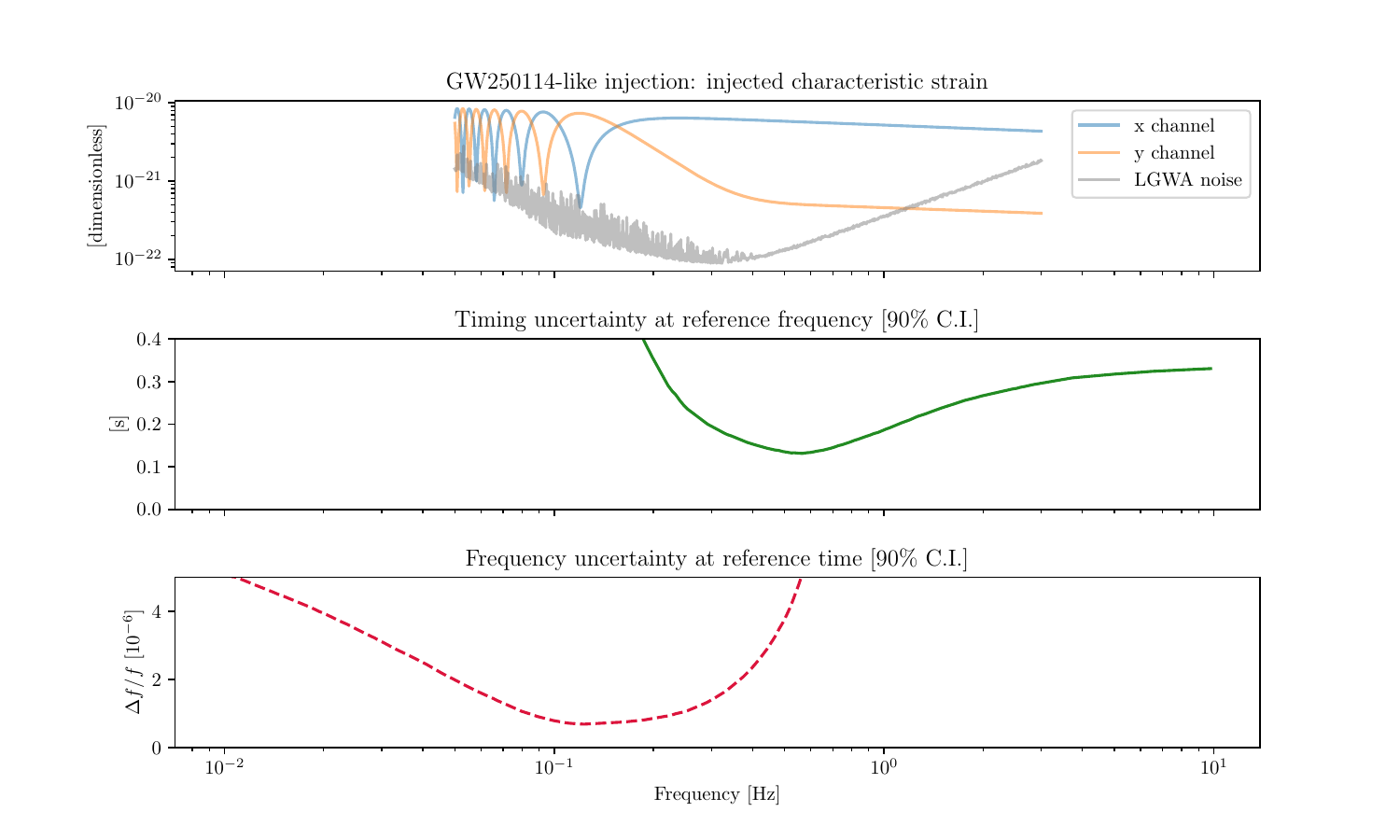}
\caption{
    Timing uncertainty as a function of frequency.
    The labels for the vertical axes are above the respective panels.
    The top panel shows the projected signal's evolution 
    compared to the LGWA noise curve. 
    The second panel shows the uncertainty in the measurement 
    of time (width of the 90\% credible interval) as a function 
    of frequency. This uncertainty also depends on the 
    reference position in the Solar System: for each frequency,
    we show the uncertainty in time as measured at that frequency's optimal 
    location, computed through a minimization procedure.
    The third panel shows the uncertainty in the measurement of 
    frequency at a reference time.
    This uncertainty is shown as the width of the 90\% credible interval
    normalized to the reference frequency.
}
\label{fig:timing-by-frequency}
\end{figure}

Figure \ref{fig:mchirp-time} shows the shape of the bivariate posterior distribution for chirp mass and time for three different choices of reference evolutionary stage: from left to right, 0.40 Hz, \MinimumUncertaintyFrequency\ and the merger; the reference frame origin is the same location (the global optimum) for all three panels.
The uncertainty in time changes, attaining a minimum at \MinimumUncertaintyFrequency\ and increasing in either direction.

The correlation structure of the distribution also changes: chirp mass is anticorrelated to time at merger, but positively correlated to time measured at a lower frequency. 
The method by which the uncertainty on time can be reduced in post-processing is by resolving this degeneracy, as well as that with other intrinsic parameters such as mass ratio and effective spin. 
Figure \ref{fig:q-time} shows this for mass ratio and time: the correlation structure is roughly opposite to that with chirp mass.
We believe this to be the reason why the optimum does not precisely achieve zero correlation with all intrinsic parameters.

Figure \ref{fig:timing-by-frequency} shows a more complete picture, as a function of frequency. For reference, in the first panel we report the characteristic strain and detector noise, similarly to Figure \ref{fig:ppd-full-bandwidth}.
In the second panel we show the width of the 90\% interval in time measured at a given frequency. 
Its minimum lies at \MinimumUncertaintyFrequency; it increases for higher frequencies, approaching an asymptote which corresponds to the uncertainty on merger time, while it quickly increases for lower frequencies.
In particular, the timing uncertatinty at the lower edge of the detector's band is quite large (see also table \ref{tab:timing-uncertainty}).
We extend the limits of this Figure beyond the detector's band to illustrate how this observation is able to provide tight constraints on the time and frequency of the signal even beyond what it directly measured. 
In a realistic context, these would be extrapolations made under the assumption that the waveform model at hand provides an adequate description of the signal.

Some works performed low-frequency parameter estimation with \ac{lisa} by parameterizing waveforms with frequency at a reference time, as opposed to time at a reference frequency \cite{toubianaParameterEstimationStellarmass2020, toubianaDetectabilityParameterEstimation2022}. 
In the third panel of Figure \ref{fig:timing-by-frequency}, we compare this approach to ours.
They use the start of observation as their reference time, but this is not a necessary choice: 
for any given frequency \(f_0\), we compute the corresponding time for the injection, \(t_{22}(f_0; \theta_0)\) and use it as a reference. 
Then, given a sample \(\theta\) in our posterior, we determine its frequency \(f\) at the time \(t_{22}(f_0; \theta_0)\).
Practically, for this step we need to numerically invert the \(t(f)\) mapping.
This yields a distribution of frequency values \(f\) at this reference time; we compute the 90\% width of this distribution and show its evolution. 
For clarity we normalize this uncertainty, \(\Delta f\),  by the injected frequency at the reference time, \(f_0\).
We used the injected waveform to find a correspondence between fiducial time and frequency, but this is only for the purposes of placing all quantities on the same axis. 
This approach is fully general and not restricted to injections. 

The optimal point differs: compared to the frequency with the highest \ac{snr} accumulation, around 0.35 Hz, the optimum for the time-at-reference-frequency approach is at a slightly higher frequency, while for the frequency-at-reference-time approach is at a slightly lower one.
We note, however, that this optimum does not correspond to the start of observation, and a frequency within the observation band is preferred.

\begin{figure}[p!]
\centering
\includegraphics[width=1.2\columnwidth]{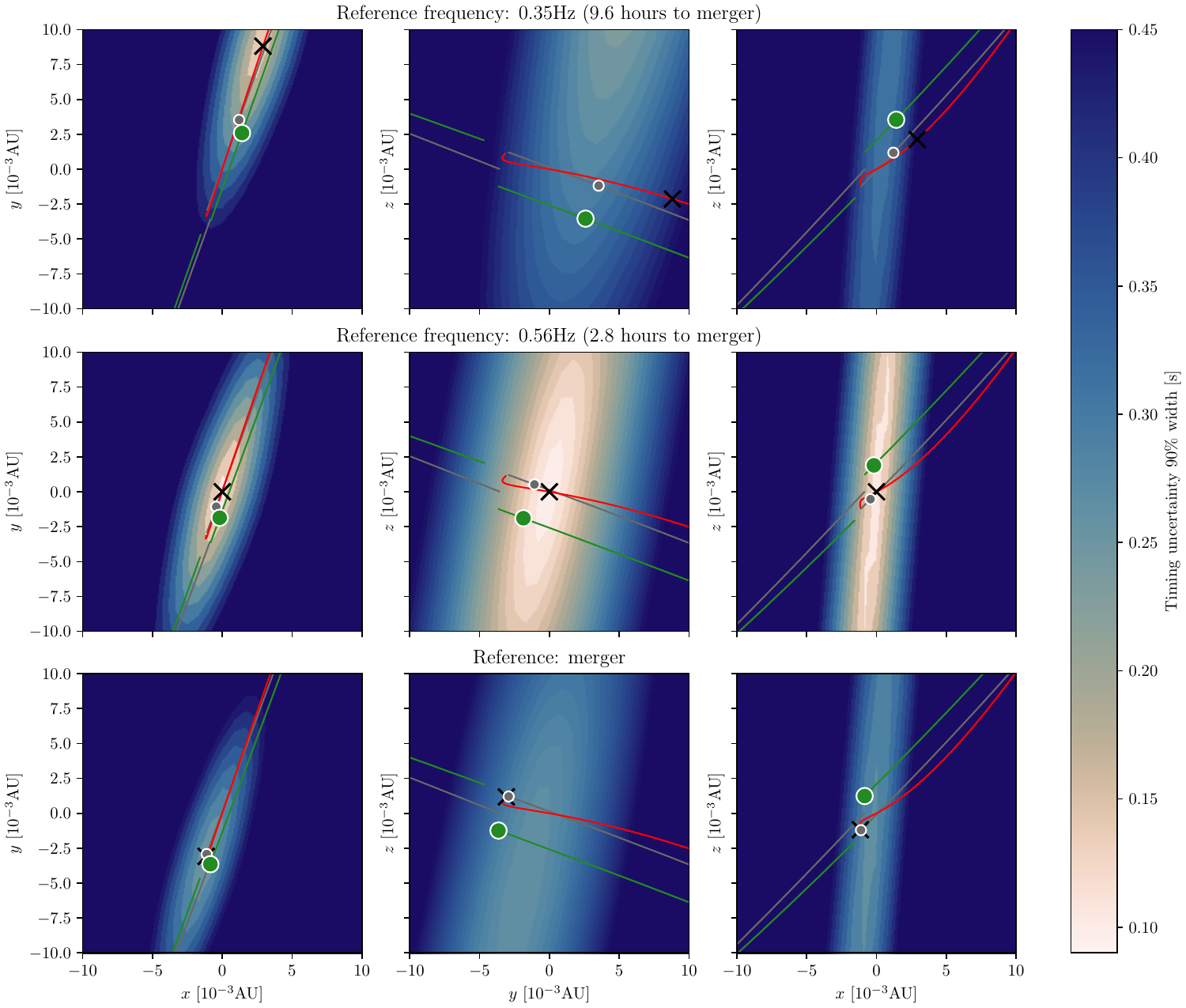}
\caption{
    Timing uncertainty as a function of position. 
    Each row shows with color the uncertainty in the measure of time at different stages, as a function of position in the Solar System: the frequency of the \(\ell |m| = 22\) mode of the waveform equalling a reference value, or the merger.
    The three columns show projections along three orthogonal axes, chosen such that 
    the \(z\) axis is along the propagation direction of the gravitational wave.
    The reference frame used for all rows is the same, and the slices
    all pass through the global minimum in uncertainty (center row).
    The green (grey) dot and line represent the projected trajectory and position of the Earth (Moon), the red line is the projected locus of the spatial minima in uncertainty as a function of frequency.
    The black cross is the projected minimum in uncertainty at the frequency of the given panel. 
}
\label{fig:timing-by-location}
\end{figure}

\begin{figure}[ht]
\centering
\includegraphics[width=\columnwidth]{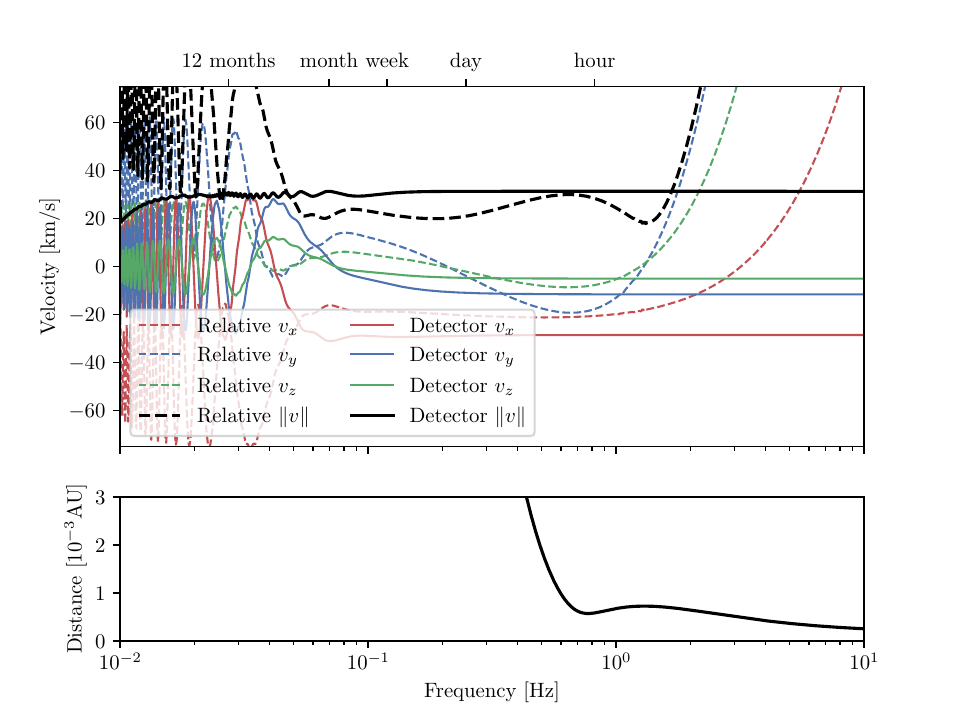}
\caption{Motion of the minimum in uncertainty, \(r_{\text{min}}(f)\), interpreted as a function of \(t_{22}(f)\). 
In the first panel we compare its velocity to that of the \ac{lgwa}, both in magnitude and component by component; 
the three axes are those of the \ac{icrs} frame. 
In the bottom panel, we show the distance between these two points as a function of frequency.}
\label{fig:minimum-uncertainty-velocity}
\end{figure}

There is an interplay between the effect of changing reference frequency and reference position: 
the optimal location at one reference frequency will in general differ from that
obtained at a different frequency.
In Figure \ref{fig:timing-by-frequency}, for each frequency we perform a minimization of the timing uncertainty as a function of position. 

Figure \ref{fig:timing-by-location} illustrates the dependence of the timing uncertainty on the position chosen in the Solar System to be the origin of the reference frame. 
In order to understand the geometry of this quantity, 
we show it in three orthogonal planes, which all pass through the global minimum in timing uncertainty. 
We choose one of the three orthogonal axes to correspond to the propagation direction of the signal, \(\hat{m}\) (Equation~\ref{eq:propagation-unit-vector}). 

For reference, we also show the trajectories and positions of the Earth and Moon, projected onto the three planes. 
The positions are computed for the time at that reference frequency, i.e.\ \(t_{22}(f)\). 

Additionally, we show the projected locus of the timing uncertainty minima: for any given frequency \(f\) and position \(r\) in the Solar System, we can use Equation \ref{eq:timing-samples-shift} to compute a set of shifted time samples \(t_i^{f, r}\). 
Then, the timing uncertainty minimum at a frequency \(f\) will be at a location \(r_{\text{min}}(f) = \text{argmin}_r \text{var}(t_i^{f, r})\), with the global minimum being \(f_{\text{min}}, r_{\text{min}} = \text{argmin}_{f, r} \text{var}(t_i^{f, r})\). 
Minimizing the variance or 90\% interval gives nearly identical results; we show the latter in Figure \ref{fig:timing-by-location} for consistency with the literature and other results in this work.

The red curve in Figure \ref{fig:timing-by-location} is the projection onto each plane of the curve \(r_{\text{min}}\); the black cross is the projection of \(r_{\text{min}}(f)\) for the reference frequency in that row.
In the first and third row, \(r_{\text{min}}(f)\) is outside of the planes, hence its apparent position outside of the timing uncertainty distribution.

As the frequency changes, \(r_{\text{min}}(f(t_{22}))\) appears to ``follow'' the Moon's position. 
We show this in more detail in Figure \ref{fig:minimum-uncertainty-velocity}: the velocity of this point approaches that of the detector near the end of the observation period. 
Indeed, their distance becomes quite small near the end of observation.
The position and motion of \(r_{\text{min}}\) is exclusively extracted from the posterior distribution: 
an approximate memory of the detectors's motion is embedded within it.

\subsection{Analytic description of timing uncertainty as a function of location} \label{sec:analytic-timing}

In Figure \ref{fig:timing-by-location}, the iso-uncertainty curves stretch significantly more along the \(\hat{m}\) direction than along the other two. 
Additionally, these curves always appear approximately elliptical.
This section will give a formal exploration of these features, as well as the consequences that can be drawn from this analysis.

Suppose that the true sky location unit vector is \(\hat{m}\). Then, we can parameterize any other sky position unit vector as \(\hat{n} = \hat{m}(1-\gamma ) + \alpha \hat{u} + \beta \hat{v}\), 
where \((\hat{m}, \hat{u}, \hat{v})\) form an orthonormal basis. These three coefficients need to satisfy \((1-\gamma )^2 + \alpha^2 + \beta^2 = 1\), since \(\lvert \hat{n}\rvert = 1\).

Let us assume there is a location \(r_0\) such that the errors on timing and sky location are uncorrelated there.
Though we did not formally prove the existence and uniqueness of this location in general, we have verified that the global minimum shown in Figure \ref{fig:timing-by-location} satisfies this condition in our case study.

For the rest of this section, we will discuss the posterior distribution for the sky position vector and timing parameter \(p(\hat{n}, t|d)\), based on which we can compute expectation values, denoted as \(\mathbb{E} \equiv \mathbb{E}_{p(\hat{n}, t|d)}\). 
We can write the sky position posterior in terms of \(\alpha\) and \(\beta\), and these can be assumed to be uncorrelated if we choose \(\hat{u}\) and \(\hat{v}\) along the principal axes of the distribution, which is always possible.
We do not explicitly write down the distribution for \(\gamma\), since it is uniquely determined by \(\alpha\) and \(\beta\). 
 
In the case of a zero-noise injection we can use a timing parameter \(t\) such that \(t=0\) at the injected value, and the mean of the \([t, \alpha, \beta ]\) vector will be \([0, 0, 0]\), while its covariance matrix will be 
\begin{align}
\Sigma = \left[\begin{array}{ccc}
\sigma_t^2 & 0 & 0 \\ 
0 & \sigma_\alpha^2 & 0 \\ 
0 & 0 & \sigma_\beta^2
\end{array}\right]
\,.
\end{align}

We want to investigate the change in this matrix when we apply the transformation
\begin{align}
\widetilde{t} = t - \frac{\Delta \vec{r}}{c}  \cdot \left(
    \hat{m}(1-\gamma ) + \alpha \hat{u} + \beta \hat{v}
\right)
\,.
\end{align}

The transformed mean will be \(\mathbb{E}([\widetilde{t}, \alpha, \beta ]) = [-(1-\bar{\gamma})\Delta \vec{r} \cdot \hat{m} / c, 0, 0]\); the new variance of the time will be:
\begin{equation}
\begin{aligned}
\mathbb{E}\left[\left(\widetilde{t}+\frac{\Delta \vec{r}\cdot \hat{m}}{c} (1-\bar{\gamma})\right)^2\right] &= \mathbb{E} \left[
    \left(
    t - \frac{\Delta \vec{r}}{c} \cdot \left(\hat{m}(\bar{\gamma}-\gamma) + \alpha \hat{u} + \beta \hat{v}\right)
    \right)^2
\right]  \\
&= \sigma_t^2
+ \left(\frac{\Delta \vec{r} \cdot \hat{u}}{c}\right)^2 \sigma_\alpha^2 
+ \left(\frac{\Delta \vec{r} \cdot \hat{v}}{c}\right)^2 \sigma_\beta^2 + \\
&\ + \left(\frac{\Delta \vec{r} \cdot \hat{m}}{c}\right)^2 \mathbb{E} \left[ (\bar{\gamma}-\gamma)^2  \right]  \\
&\ - 2 \frac{\Delta \vec{r} \cdot \hat{m}}{c}
\bigg(
    \mathbb{E} \left[t (\gamma -\bar{\gamma})  \right] \\
    &- \frac{\Delta \vec{r} \cdot \hat{u}}{c} \mathbb{E} \left[ (\gamma-\bar{\gamma}) \alpha  \right]
    - \frac{\Delta \vec{r} \cdot \hat{v}}{c} \mathbb{E} \left[ (\gamma-\bar{\gamma}) \beta   \right]
\bigg)
\,.
\end{aligned}
\end{equation}

If the sky localization for a given source is tight, we will have \(|\alpha |, |\beta | \gg |\gamma |\) for samples in the posterior bulk. 

If we neglect the terms containing \(\gamma\), we are left with a simple geometry: the uncertainty is minimal along a line parallel to \(\hat{m}\) and passing through \(\Delta \vec{r} = 0\); iso-variance contours on planes orthogonal to this line are ellipses, whose axes are determined by the uncertainty in sky position. 

We can understand the typical variation scale of these ellipses by considering variations in \(\Delta \vec{r}\) along an axis in the plane spanned by \(\hat{u}\) and \(\hat{v}\) --- this will be a linear combination of \(\alpha\) and \(\beta\) we denote as \(\theta\). We seek the value \(r _{\text{char}}\) of the radial coordinate along this axis for which the timing variance doubles compared to its minimum.
The equation to solve is 
\begin{align}
2 \sigma _t^2 = \sigma _t^2 + \frac{r_{\text{char}}^2}{c^2} \sigma^2_\theta 
\,,
\end{align}
which is solved by \(|r_{\text{char}}| = \sigma _t c / \sigma _\theta \).

This allows us to express the general heuristic that the choice of origin for our reference frame of a point within Earth's orbit will have a significant impact on the degeneracy between parameters when 
\begin{align}\label{eq:characteristic-radius-by-uncertainties}
|r_{\text{char}}| = \frac{\sigma _t c}{\sigma_{\theta} } \ll \text{AU}
\,
\end{align}
for any angular variable \(\theta\).

This is a fully general line of reasoning, whose application is not limited to lunar gravitational wave detection. In order to illustrate this, we apply it to the original detection of GW250114 by the \ac{lvk}.
For ground-based \ac{cbc} observations, Equation \ref{eq:characteristic-radius-by-uncertainties} will yield much smaller typical radii than those of the \ac{lgwa}: the timing precision is much better, typically on the millisecond scale or better, while the angular posteriors can be relatively wide, especially if two or fewer interferometers are operational. 
The uncertainty on geocentric time reported by the collaboration is \(\sigma_t \approx 0.92\text{ms}\).
As for \(\sigma_\theta\), since the distribution 
is highly elongated we will have two very different uncertainties: by performing \ac{svd} on angular coordinates defined as deviations from the maximum likelihood point, we get \(\sigma_\alpha \approx 0.97 \text{deg}\) and \(\sigma_\beta \approx 18 \text{deg}\).
The predicted uncertainty variation scales \(|r_{\text{char}}|\), based on this, are 16000km and 900km respectively, though we should note that \(\gamma\) in this case is not negligible, with a standard deviation of 7.7 degrees.

Indeed, through numerical minimization of the uncertainty we find that at the position \([-3.25, 3.39,  4.01] \times 10^6 \text{m}\), in \ac{icrs} coordinates from the center of the Earth, the \(1\sigma\) timing uncertainty for GW250114 as observed by the \ac{lvk} was only \(71\text{\textmu s}\): more than an order of magnitude reduction from the geocentric frame.

\subsection{Timing parameterization impact on sampling time} \label{sec:sampling-time}

Using a reference frame where the timing uncertainty is much higher than its minimum value, 
despite being formally equivalent to any other reference frame, has a significant impact on 
the efficiency of parameter estimation. 
This is because stochastic samplers are more efficient at exploring distributions with regular 
shapes and small correlations. 

We illustrate this with a minimal example, in which we once again analyze 
a signal compatible with GW250114, but we vary only the sky position and 
arrival time, fixing all other parameters to their injected values. 
Additionally, we use a phase-marginalized likelihood, since 
conditioning on a fixed value for the phase leads to banded structures 
in the sky distribution which we do not wish to explore.
The results, in terms of sampling time and number of likelihood evaluations, are summarized in Table \ref{tab:sampling-time-comparison}.

\begin{table}
\begin{tabular}{cccccc}
\hline
 \makecell{Likelihood \\ evaluations 
 [$\times 10^5$]
 } 
 & \makecell{Sampling \\ time [min]} 
 & \makecell{Timing \\ uncertainty [s]} 
 & \makecell{Prior \\ width [s]} & Sampler & Origin \\
\hline
\rowcolor{green!15} 0.9 & 3.0 & 0.075 & 0.5 & nessai & optimal \\
\rowcolor{green!5} 1.1 & 3.6 & 0.075 & 27.0 & nessai & optimal \\
\rowcolor{red!15} 6.6 & 16.7 & 4.733 & 27.0 & nessai & SSB \\
\rowcolor{green!15} 8.1 & 16.8 & 0.074 & 0.5 & dynesty & optimal \\
\rowcolor{green!5} 15.3 & 33.0 & 0.076 & 27.0 & dynesty & optimal \\
\rowcolor{red!15} 91.4 & 198.6 & 4.766 & 27.0 & dynesty & SSB \\
\hline
\end{tabular}
\caption{Summary of the computational cost for the three-parameter example described in \ref{sec:sampling-time}.}
\label{tab:sampling-time-comparison}
\end{table}

We compare two different reference frames: one with origin at the SSB, 
and one with origin at the point which minimizes timing uncertainty for this problem.
Since we are conditioning on several parameters, 
the minimum here is slightly different from that found in \ref{sec:shifting-timing}.
Furthermore, since all intrinsic parameters are fixed it is equivalent 
to use the time at merger or at any other frequency as a reference.

Since the uncertainty on time at the SSB is much higher than that at the optimum, 
a wider prior is required for an unbiased analysis. 
We choose the prior widths to approximately correspond 
to 6 times the 90\% posterior width in both cases.

Sampling is faster in the optimal reference frame
because we are able to use a narrower prior,
but also because the structure of the likelihood becomes simpler. 
In order to distinguish these two contributions, 
we include a scenario where we use the same prior width 
as that which is needed in the SSB frame, 
but we use the optimal reference. 

Furthermore, we compare two different, commonly used samplers.
In both cases the underlying algorithm is Nested Sampling, 
with the same number of live points (2000).
The difference lies in the way new live points are sampled 
from the likelihood-constrained prior:
with \texttt{dynesty} \cite{speagleDYNESTYDynamicNested2020} 
this is done by estimating a bounding region for the current 
live points; in the default setting we use this region is a set of ellipsoids. 
With \texttt{nessai} \cite{williamsNestedSamplingNormalising2021}, 
a normalising flow is trained on the samples, 
which leads to significantly more efficient proposals. 

Overall, we see that \texttt{nessai} requires around an order of magnitude fewer likelihood evaluations than \texttt{dynesty}, and correspondingly a lower sampling time. 
With both algorithms, however, 
using a suboptimal reference frame origin 
leads to significantly increased sampling time 
and likelihood evaluations. 
The increase is approximately a factor 5 with \texttt{nessai} and a factor 10 with \texttt{dynesty}; 
we attribute this to the fact that the former algorithm 
is more flexible, and therefore better able to model
the degenerate shape of the distribution. 
Nevertheless, using the optimal reference frame 
origin incurs no additional computational cost, 
and there is no reason not to adopt it.

\subsection{Sky localization as a function of orbital area} \label{sec:localization-function-area}

We wish to investigate the dependence of the sky localization on the geometry of the detector's orbit, \emph{i.e.} the inverse-area scaling suggested by Equation \ref{eq:wen-chen-area}. 

The statistical area introduced in Equation~\ref{eq:wen-chen-area} is defined as
\begin{equation}\label{eq:statistica_area}
    A_s(T)= 2\pi \sqrt{\mathrm{Var}(r_\theta)\mathrm{Var}(r_\phi)-\mathrm{Cov}^2(r_\theta,r_\phi)}\, ,
\end{equation}
where Var and Cov are variance and co-variance with expectation values weighted with respect to the accumulation of SNR, and $r_\theta \equiv \partial_\theta \mathbf{n} \cdot \mathbf{r}(t)$ being the projection of the detector motion vector. We approximate the weighting function in the frequency domain as follows (for further details see Appendix~\ref{subsec:area}):
\begin{equation}\label{eq:weight_xidot}
w(t) = \left|\frac{\mathrm{d} f}{\mathrm{d}t}\right| \frac{\left|h_x(f)\right|^2  + \left|h_y(f)\right|^2}{S_n(f)} \,.
\end{equation}

\begin{figure}[ht]
\centering
\includegraphics[width=\defaultwidth]{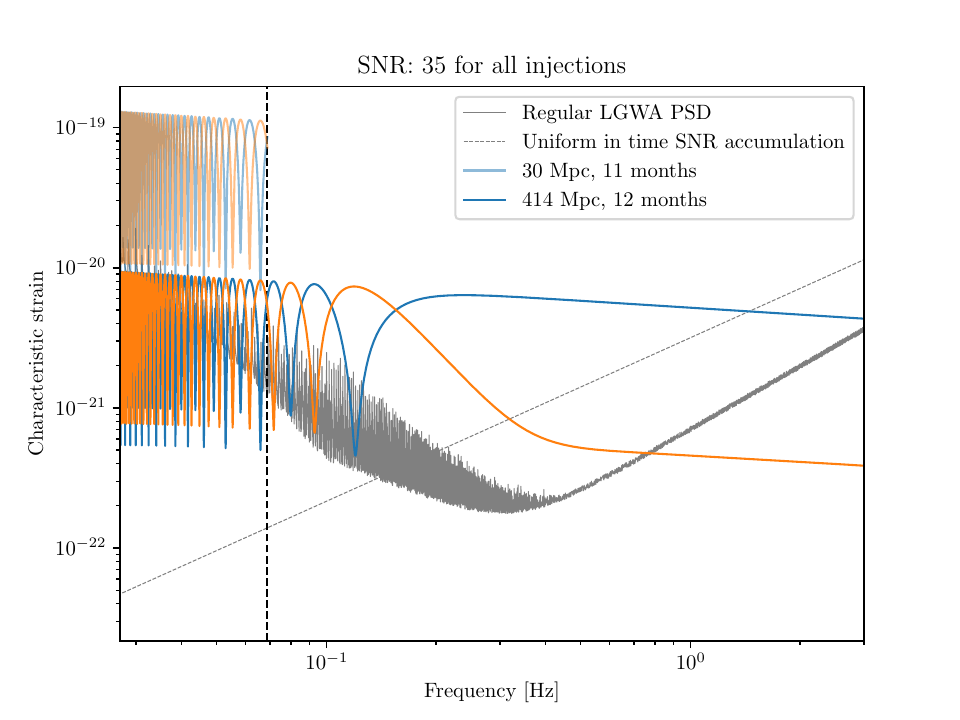}
\caption{Characteristic signal strain (\(2 f |h(f)|\)) for the 12-month and 11-month injections, as well characteristic noise strains (\(\sqrt{f S_n(f)}\)) corresponding to the \acp{psd} used for in the injection scenarios in section \ref{sec:localization-function-area}.
For both injections, we show in blue and orange the strains for the \(x\) and \(y\) \ac{lgwa} channels respectively.
The other injections, with durations of 1 to 10 months, are scaled analogously.}
\label{fig:waveforms-month-by-month}
\end{figure}

To explore the dependence of the sky localization on the area, we set up 12 identical injections with durations ranging from 1 to 12 months, with the same starting time and frequency but differing ends.
All else being equal, the \ac{snr} would be significantly different; to compensate for this effect, we adjust the distance for each injection so that the \ac{snr} is constant.
This approach allows us to isolate the geometric effects from Equation \ref{eq:wen-chen-area}.
All injections except for the 12 month one are unrealistically close, with distances of 30Mpc or less.
In Figure \ref{fig:waveforms-month-by-month} we show the projected waveforms for the 11 and 12 month case for reference; the remaining ones are analogous, with higher and higher amplitudes, and the same starting frequency of 27 mHz.

To show the impact of the SNR weighting applied when computing \(A_s\), we compare the Fisher prediction to one computed using the \textit{geometrical} area $A_g$.

\begin{figure}[ht]
\centering
\includegraphics[width=\columnwidth]{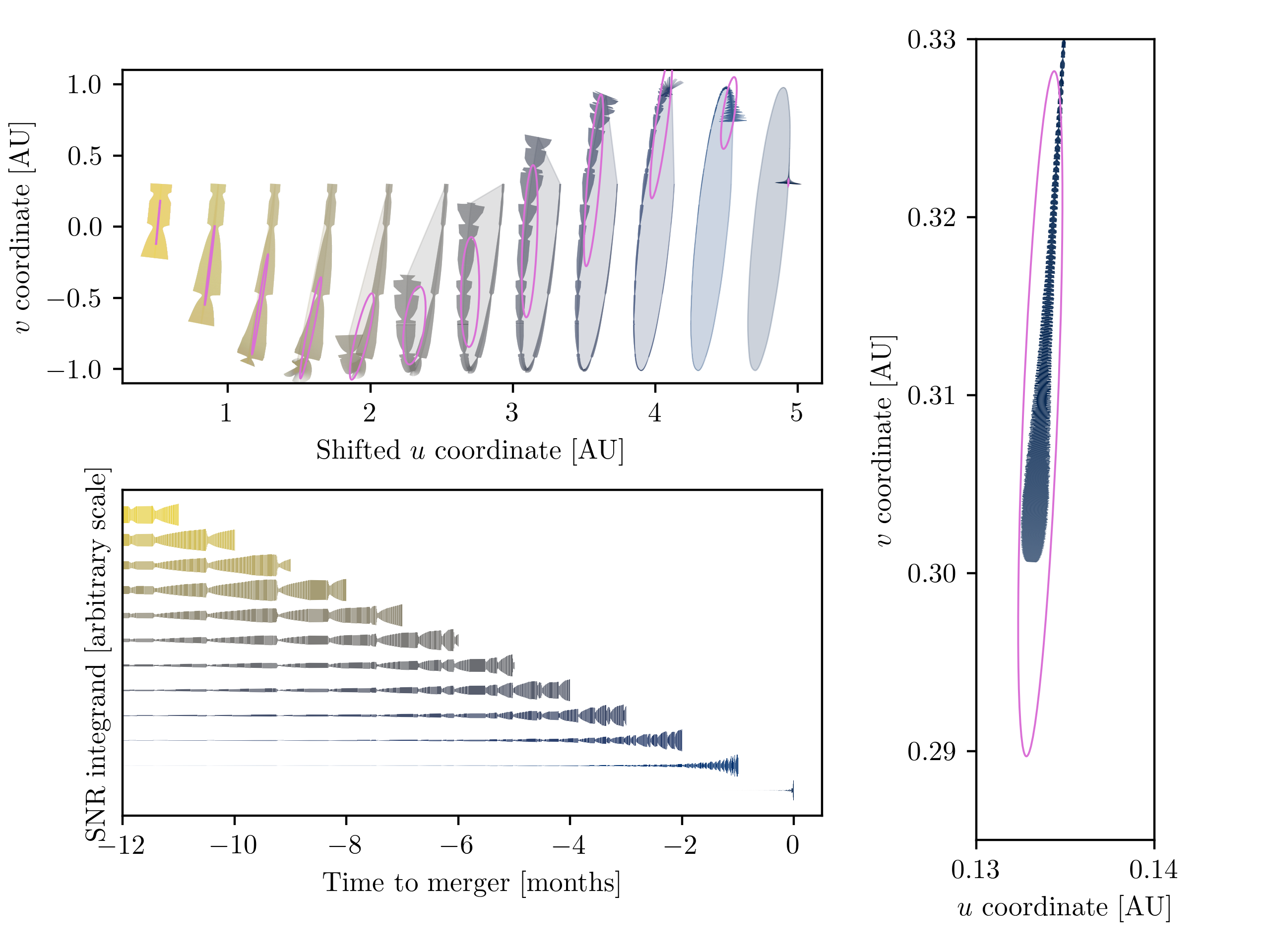}
\caption{Top left panel: trajectories and spanned areas for the 12 injections. They are projected onto a plane orthogonal to the injected GW's propagation direction \(m\). The \ac{ssb}'s position is shown as a dot. Colors ranging from blue to yellow are associated with the length of the injection, analogously to Figures \ref{fig:sky-localization-areas} and \ref{fig:sky-localization-vs-theory}.
For each trajectory, we also show the accumulation of SNR through the weight function \(w(t)\), defined by Equation~\ref{eq:weight_xidot}, whose root-mean-square computed over time is the total \ac{snr}.
Furthermore, we shade the geometric areas \(A_g\), and show in pink the ellipses corresponding to \(A_s\). 
Bottom left panel: the quantity \(w(t)\) as a function of time for the same injections. In all cases, we normalize \(w(t)\) so that its maximum value is fixed.
Right panel: a detailed view of the 12-month trajectory shown in the top left panel.}
\label{fig:trajectories-month-by-month}
\end{figure}

In order to precisely compute this area, we project the Moon's motion onto a plane orthogonal to the propagation direction of the \ac{gw}.
Then, the 2D area of the convex hull of the points gives us \(A_g(T)\), while the distribution of the positions weighted as described in Eq.~\ref{eq:weight_xidot} gives us \(A_s(T)\).
We show both areas in Fig~.\ref{fig:trajectories-month-by-month}. For more details on the scaling of these areas, see Sec.~\ref{fig:areas-scaling} in the Appendix.

We perform the injections in three different configurations in order to highlight the impact of different aspects:

\begin{enumerate}
    \item The \textbf{all parameters} scenario: these are zero-noise injections performed recovering all parameters. For computational efficiency we do not consider spin, which is fixed to its fiducial value --- we verified that this assumption has little effect on the sky posteriors.
    \item The \textbf{simplified} scenario: here, we only vary sky position, time, and phase. 
    \item The \textbf{simplified, uniform SNR} scenario: the parameters are the same as the simplified one, but the detector \ac{psd} is changed. Instead of the \ac{lgwa} \ac{psd}, we set \(S(f) \propto f^{4/3}\),\footnote{The integrand to consider to this end is 
\begin{align}
  \frac{|h(f)|^2}{S_n(f)} \dd{f} = w(t) \dd{t} \propto \frac{f^{-14/6}}{f^{\kappa}} \frac{\dd{f}}{\dd{t}} \dd{t} 
\,,
\end{align}
where \(t(f) \propto f^{-8/3}\), while at leading order \(|h(f)| \propto f^{-7/6}\). Due to the Moon's rotation a modulation also affects the projected signal, but the \(f^{-7/6}\) scaling is correct when averaging on a month-by-month basis.
By imposing the condition of the integrand in time being constant as a function of time we get \(\kappa = -14/6 + 11/3 = 4/3\).
This \ac{psd} is shown with a dashed line in Figure \ref{fig:waveforms-month-by-month}. } normalized such that the \ac{snr} is the same as all the other injections. 
\end{enumerate}

\begin{figure}[ht]
\centering
\includegraphics[width=\columnwidth]{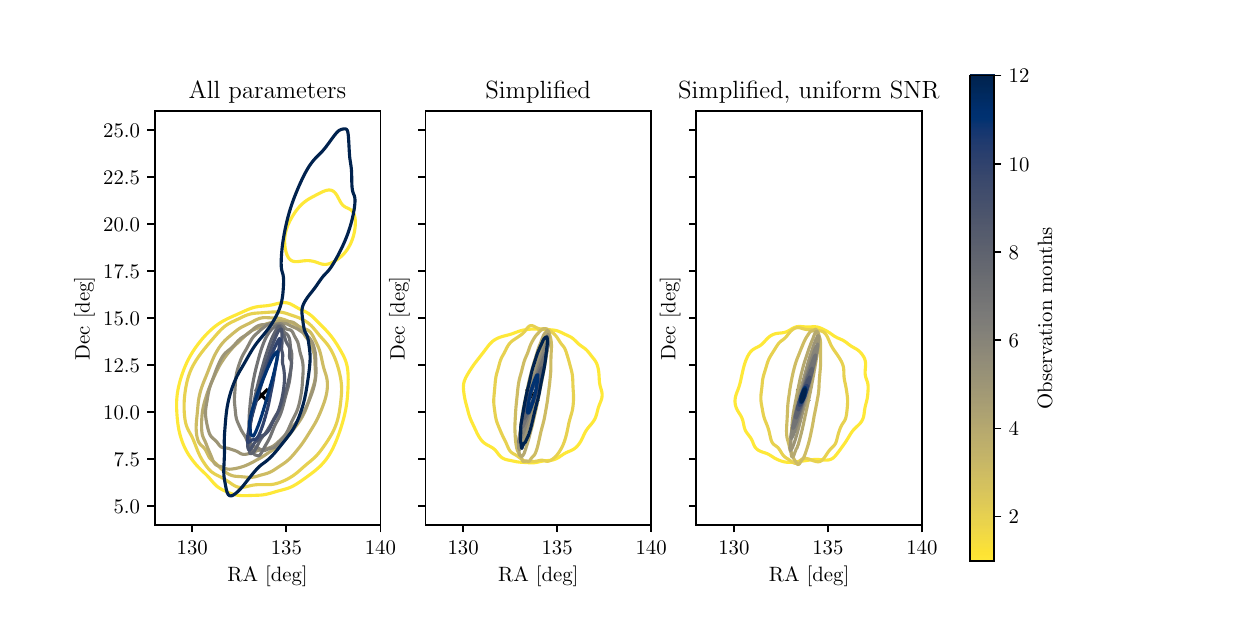}
\caption{90\% posterior credible areas for the 12 injections considered in the three scenarios of Sec.~\ref{sec:localization-function-area}.}
\label{fig:sky-localization-areas}
\end{figure}

While the SNR is constant throughout these injections by construction, the sky area varies: we show the 90\% sky localization contours in Figure \ref{fig:sky-localization-areas}.

\begin{figure}[ht]
\centering
\includegraphics[width=\defaultwidth]{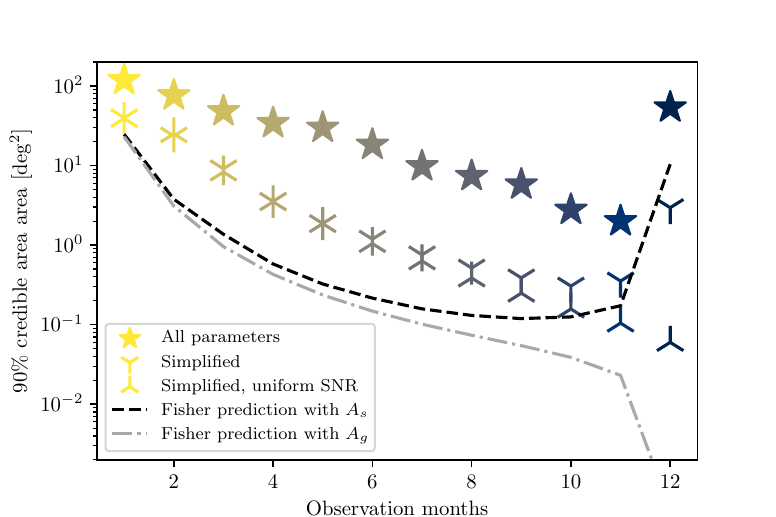}
\caption{Comparison between the areas obtained through parameter estimation and the theoretical prediction of Equation \ref{eq:wen-chen-area}, shown as a dashed black line; we also compute an analogous prediction replacing the statistical area \(A_s\) with the geometrical one, \(A_g\), and show it with a dash-dotted grey line. The SNR is fixed for all injections.}
\label{fig:sky-localization-vs-theory}
\end{figure}


Figure \ref{fig:sky-localization-vs-theory} summarizes the results from all injections. 
We compare them with the prediction of Equation \ref{eq:wen-chen-area}, computing the value of \(f^2 \rho_T^2\) by evaluating the scalar product \(\langle f h | f h\rangle\) (see Equation~\ref{eq:innProd}). 

The scaling behavior for the sky localization areas is qualitatively consistent with the prediction of Equation \ref{eq:wen-chen-area} for all injections lasting between 1 and 11 months.
The 12-month case, in which the frequency reaches the upper edge of the detector band, is significantly different: the area is significantly larger than the 11-month case, while Equation \ref{eq:wen-chen-area} would predict it to be smaller.
This applies both for the all parameters scenario and the simplified one: it is due to the fact that most of the \ac{snr} is accumulated in the last month of observation, during which the curvature of the lunar orbit is not significant. 
Indeed, if we counteract this effect by artificially altering the detector \ac{psd} and imposing that every month accumulates the same amount of \ac{snr} (the ``simplified, uniform \ac{snr}'' scenario), the decrease in area continues for the 12-month injection.

The bimodality observed in the full injection is only present in the 12-month case with all parameters; with a shorter observation the localization is driven by phase as opposed to amplitude modulations, while in both simplified scenarios we condition on the correct value of the inclination angle, so the degeneracy is artificially removed.


Overall, the qualitative scaling behaviour is indeed observed, but the formula fails to provide a quantitatively correct estimate, and generally underestimates the localization area, especially when considering a realistic set of parameters. 
Indeed, the analytical expression of Equation~\ref{eq:wen-chen-area} is obtained assuming the signal to be exactly known except for the signal time and the sky localization, which introduces a discrepancy with the realistic case (see Appendix~\ref{subsec:area}). 

\section{Conclusions} \label{sec:conclusions}

The most common reference frame origins for \ac{gw} analyses are the geocenter and the \ac{ssb}. The former is adequate for current \ac{cbc} analyses with ground-based detectors, though the Earth's orbital motion may become relevant for very low-mass signals and/or future detectors \cite{chenFastFrequencydomainExpression2024}. The latter is generally used for the analysis of long signals, both with space- and ground-based detectors.

In this work, we have shown that the precision and efficiency of parameter estimation for long-duration GW signals in the deci-Hertz band are fundamentally linked to the geometry of the detection. Using a frame comoving with the SSB, but with origin elsewhere, can lead to a significant acceleration of parameter estimation, especially for signals observed with high timing precision but comparatively low sky localization precision. This is the case for stellar-mass BBH signals observed by the LGWA: they last for a large fraction of a year in band, but accumulate most of their SNR in the final part of the observation, at high frequency and in a small region within the Solar System. 

Furthermore, we find that merger time is a suboptimal parameter choice for the \ac{lgwa}, since it happens outside the detector's band. We explore two alternative choices (time at a reference frequency, or vice versa), both of which show a global minimum in uncertainty as a function of the chosen reference time or frequency, close to the minimum of the detector's PSD (Figure~\ref{fig:timing-by-frequency}). At these global minima, the degeneracies between the timing parameter and the source's intrinsic parameters are small. 
However, the reduction in timing uncertainty is only by a factor 2--3 compared to the merger time parameter (Tab.~\ref{tab:timing-uncertainty}). 
The four reference parameters we identified---reference frequency and three coordinates for the position in the solar system---can be varied in post-processing, yielding equivalent analyses. There is a global optimum in this four-dimensional space which minimises the uncertainty on the timing parameter; it can be found through a preliminary analysis and employed thereafter. At this optimum, the correlations between timing, intrinsic parameters and sky position are minimised, which can significantly accelerate parameter estimation.
We perform a simplified experiment (Sec.~\ref{sec:sampling-time}) to quantify the relation between timing uncertainty and inference performance: we observe an acceleration of inference cost by a factor 5--10 (depending on the sampler) when the timing uncertainty was reduced by a factor 63.

We validate our findings with GW250114 \cite{abacGW250114TestingHawkings2025} as a case study.
If the \ac{lgwa} had been operational in the year 2024 and early 2025,
it would have placed significantly tighter constraints on some of its parameters, like chirp mass and sky position, as well as providing several days of early warning, despite observing this source with a lower \ac{snr} than the \ac{lvk} collaboration's result.

We compare our results with an analysis of the same injection with the Einstein Telescope in two of its proposed configurations. 
The parameter constraints it provides are complementary to those of the \ac{lgwa}: for example, their posterior distributions are approximately 
orthogonal in \((\mathcal{M}, q)\) space, making their intersection quite narrow; furthermore, the \ac{lgwa} posterior distribution on the 
sky position would exclude one of the two modes we observe for the \ac{et}-\(\Delta\) configuration.
Notably the precision on certain parameters, such as the chirp mass, is better with the \ac{lgwa} than with \ac{et} in either of its configurations, despite a lower \ac{snr} by more than an order of magnitude.
This hints at the possibility for such low-frequency observations to also provide exceptional constraints on other effects on the early inspiral, such as eccentricity or matter effects.

The gold standard for this type of combined analysis is to preserve phase coherence across detectors \cite{wuMultibandParameterEstimation2025};
we hope to explore this for lunar gravitational wave detection in future work.

We make some simplifying assumptions, such as a known lunar response, and an absence of data gaps.
A more realistic consideration of these is likely to negatively impact the analysis; 
however, we expect that since these are mainly amplitude corrections, 
while the parameter constraining ability of the \ac{lgwa}
is mostly driven by phase, the global picture 
we obtain here should be robust.
We expect our other assumptions, such as working in zero-noise, only using the \(\ell |m|=22\) mode without precessing spin and eccentricity, will either have little impact on the posterior width, or be approximately satisfied for real signals. 

Finally, we connect the ability of the \ac{lgwa} to localize to the analytical Fisher-matrix scaling predicted in reference \cite{wenGeometricalExpressionAngular2010}, and found it to be qualitatively accurate, though imprecise in its normalization when compared to a full Bayesian analysis.
Specifically, we confirm that the sky localization area is inversely proportional to the weighted area described by the detector's orbit, \(A_s\) (Eq.~\ref{eq:statistica_area}). 
In the case of GW250114, this is significantly smaller than the full geometric area described by the detector's orbit, \(\pi R_*^2\). 
For less massive binaries, \(A_s\) will cover a larger fraction of the geometric area, as illustrated in Fig.~\ref{fig:areas-scaling}. 
The area \(A_s\) is defined on the orthogonal plane to the \ac{gw}'s propagation direction. 

This scaling has another important implication:
for a realistic space mission, it is plausible that 
data will be unusable in certain windows.
For the \ac{lgwa}, this is expected to happen with relative frequency, on the order of one hour every day, 
due to the long ringing of deep Moonquakes \cite{steinIntroductionSeismologyEarthquakes2009,ajithLunarGravitationalwaveAntenna2025}. 
This poses a data analysis challenge \cite{burkeMindGapAddressing2025}, but through the 
scaling in Equation \ref{eq:wen-chen-area} we can 
confidently state that relatively short missing 
data chunks of this kind will only 
impact sky localization by slightly decreasing the total \ac{snr}: 
the area described by the detector's orbit 
is nearly unchanged if a small sliver of the perimeter
is removed, since the trajectory is nearly linear on those scales.

Our results relating to the localization accuracy of a moving detector and timing parameterization are applicable to any long \ac{cbc} observation, not necessarily from the Moon.
Our work shows that inference for long-duration GW signals is 
fundamentally a geometric problem, and that exploiting this structure is 
essential for both accuracy and efficiency.

\paragraph{Software and data availability}

The \ac{lgwa} response and likelihood were modelled with the \texttt{lgwa-response} package, \url{https://github.com/jacopok/lgwa-response}.
The code for the injections and post-processing, as well as posterior samples from all the injections, can be found in the companion Zenodo data release \cite{data_release}.

This work also made use of the following software packages: \texttt{astropy} \cite{astropy:2013,astropy:2018,astropy:2022,astropy_17756022}, \texttt{matplotlib} \cite{Hunter:2007}, \texttt{numpy} \cite{numpy}, \texttt{python} \cite{python}, \texttt{scipy} \cite{2020SciPy-NMeth,scipy_18736568}, \texttt{Bilby} \cite{bilby_paper,bilby_paper_2,Bilby_18788906}, \texttt{corner.py} \cite{corner-Foreman-Mackey-2016,corner.py_14209694}, \texttt{nessai} \cite{williamsNestedSamplingNormalising2021} and \texttt{Numba} \cite{numba:2015,Numba_19720971}.

Software citation information was aggregated using \texttt{\href{https://www.tomwagg.com/software-citation-station/}{The Software Citation Station}} \cite{software-citation-station-paper,software-citation-station-zenodo}.

\paragraph{Acknowledgements}

J.T.~is grateful to Michael Williams for fruitful discussions related to \texttt{nessai}. 
J.T.~acknowledges funding from the Della Riccia Foundation under an Early Career Scientist Fellowship.
This work is supported by the Italian Space Agency (ASI) under Grant No.~2025-29-HH.0.
F.I.~is supported by a Miller Postdoctoral Fellowship and by NSF Grants No.~AST-2307146, No.~PHY-2513337, No.~PHY-090003, and No.~PHY-20043, by NASA Grant No.~21-ATP21-0010, by John Templeton Foundation Grant No.~62840, by the Simons Foundation [MPS-SIP-00001698, E.B.], by the Simons Foundation International [SFI-MPS-BH-00012593-02], and by Italian Ministry of Foreign Affairs and International Cooperation Grant No.~PGR01167.
F.S.~has been funded by the European Union – NextGenerationEU under the Italian Ministry of University and Research (MUR) "Decreto per l'assunzione di ricercatori internazionali post-dottorato PNRR" - Missione 4 "Istruzione e Ricerca" Componente 2 "Dalla Ricerca all'Impresa" del PNRR - Investimento 1.2 “Finanziamento di progetti presentati da giovani ricercatori” - CUP D13C25000700001. 

\bibliography{refs20260603}

\appendix

\section{Injection details} \label{app:injection-details}

\begin{figure}[ht]
\centering
\includegraphics[width=\columnwidth]{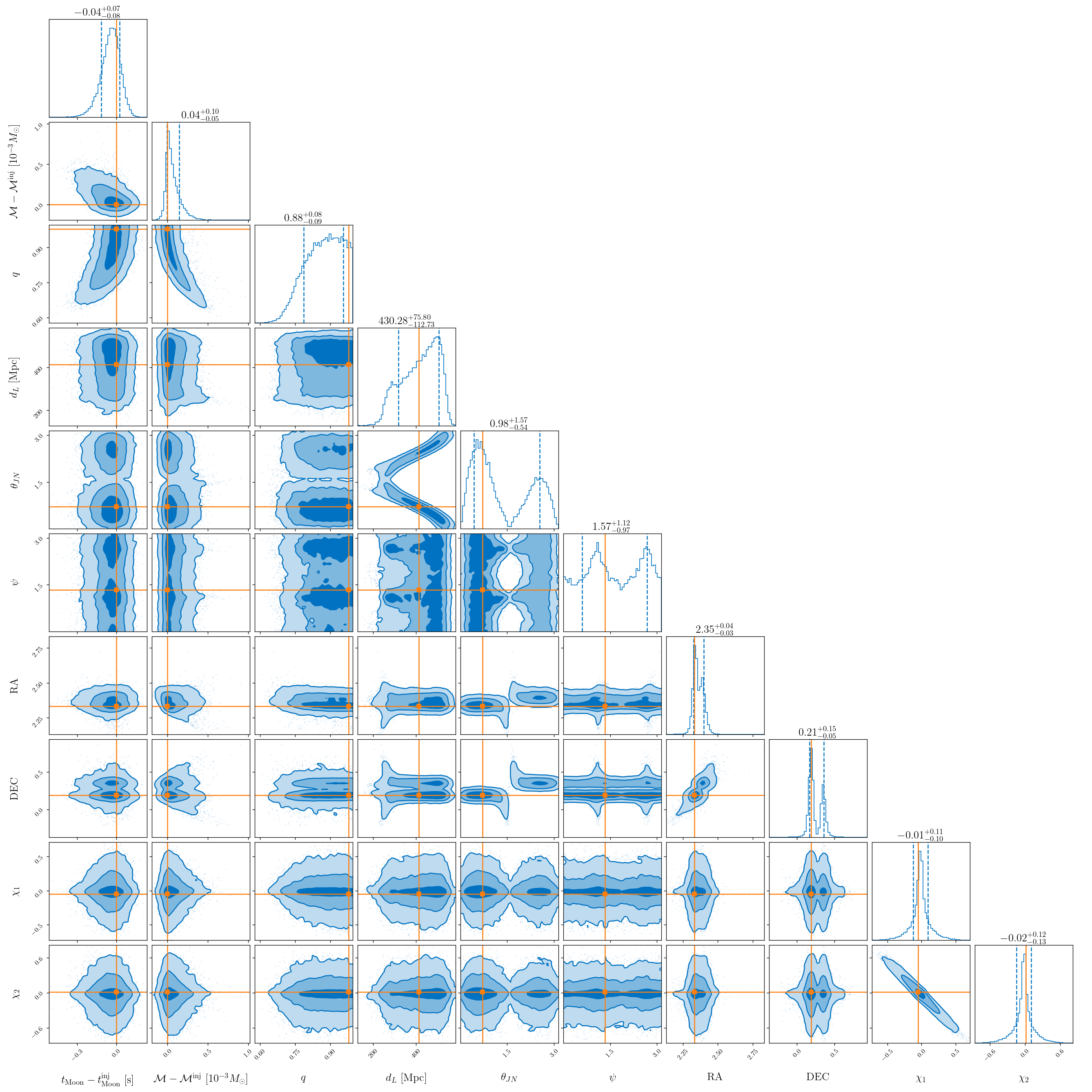}
\caption{Corner plot for our injection of GW250114 with the LGWA. Due to the large ratio between value and uncertainty, for the timing parameter and chirp mass we show the distribution of the difference between the inferred and injected values.}
\label{fig:corner-full-bandwidth}
\end{figure}

In Figure \ref{fig:corner-full-bandwidth} we show a corner plot including all sampled parameters for our GW250114-like injection. In table \ref{tab:injection-parameters} we report the injected values of the parameters.

\begin{table}
\begin{tabular}{cccc}
\hline
Parameter & Symbol & Injected value & Unit\\
\hline
Chirp mass & $\mathcal{M}$ & 31.27178 & $M_\odot$ \\
Mass ratio & $q$ & 0.98 &  \\
Luminosity distance & $d_L$ & 414 & Mpc \\
Inclination angle & $\theta_{JN}$ & 0.72 & rad \\
Polarization angle & $\psi$ & 1.33 & rad \\
Reference phase & $\phi$ & 1.57 & rad \\
Right ascension & \texttt{ra} & 2.3 & rad \\
Declination & \texttt{dec} & 0.19 & rad \\
Time at Moon & $t_{\mathrm{Moon}}$ & 1420878140.0 & s \\
Primary aligned spin & $\chi_1$ & -0.05 &  \\
Secondary aligned spin & $\chi_2$ & 0.01 &  \\
\hline
\end{tabular}
\caption{Injected parameters, corresponding to the maximum likelihood point in the analysis by the \ac{lvk} collaboration \cite{abacGW250114TestingHawkings2025}. The injected time for the signal's arrival at the Moon is computed as \(t _{\text{geo}} + \Delta t\), with \(\Delta t = (\vec{r}_{\text{Earth}} - \vec{r}_{\text{LGWA}}) \cdot \hat{m} / c \approx -1.2 \text{ s}\), where \(\hat{m}\) is the sky direction vector of the \ac{gw} signal as computed with the injected parameters, \(c\) is the speed of light and the positions of the Earth and the \ac{lgwa} detector are computed at the time 
\(t_{\text{geo}}\). This involves the small approximation of neglecting the Moon's motion during the 1.2 s delay between the two points.}
\label{tab:injection-parameters}
\end{table}

We use the \texttt{nessai} sampler, with the following settings:  \texttt{nlive}: \texttt{5000}, \texttt{stopping}: \texttt{0.05}, \texttt{reset\_flow}: \texttt{20}, \texttt{volume\_fraction}: \texttt{0.97}.
These are similar or more conservative than those typically used in \ac{gw} analyses.

We use the following configuration for the normalizing flow: \texttt{flow\_proposal\_class}: \texttt{"gwflowproposal"}, \texttt{n\_blocks}: \texttt{10}, \texttt{n\_layers}: \texttt{4}, \texttt{n\_neurons}: \texttt{48}, and for the training configuration \texttt{patience}: \texttt{25},
and \texttt{max\_epochs}: \texttt{1000}.
The network's size is larger than those typically used in \ac{gw} analyses, and an allowance is given for slower training. 
We have found by trial and error that this improves overall performance, since the proposal efficiencies of smaller networks are significantly worse.

\subsection{Relative binning convergence} \label{app:relative-binning}

The full frequency-domain likelihood for 1 year of data, sampled at 6 Hz to achieve a Nyquist frequency of 3 Hz, would require us to use a frequency grid of nearly 20 billion points for each \ac{lgwa} channel.
To accelarate this we employ relative binning \cite{zackayRelativeBinningFast2018}, which allows for a significantly sparser frequency grid by employing a reference waveform and writing integrals in terms of the ratio of any given waveform to this reference. 
This technique can be generalized to include higher order modes \cite{narolaRelativeBinningComplete2023}, though the single-mode version from the original paper suffices for our purposes.

We fix the reference waveform \(h_0 = h(f; \theta_0)\) to be the injected one, and in order to compute the likelihood at a generic point \(\theta\) in parameter space we use the ratio \(r(f) = h(f; \theta ) / h_0(f)\).
This quantity exhibits a much slower variation as a function of frequency than the waveform itself; however, it does still oscillate due to the rotation of the Moon: it is important that the frequency grid is fine enough to resolve this.

Since these oscillations happen at low frequencies, we expect the best frequency grid to densely sample that region. 
We investigate this by computing the posterior-weighted distribution of the log-likelihood error for different grid choices.

We can obtain generic ``polynomial'' grids with an exponent \(\alpha\) in the band \([f_{\text{min}}, f_{\text{max}}]\) by computing uniformly-spaced samples \(\widetilde{f}_i\) between \(f_{\text{min}}^{\alpha}\) and \(f_{\text{max}}^{\alpha}\), and then applying the transformation \(f_i = \widetilde{f}_i^{1/\alpha}\).
The \(\alpha = 1\) case corresponds to a uniform grid in frequency. 
The \(\alpha = -8/3\) case corresponds to a uniform grid in time (to the lowest post-Newtonian order).
The \(\alpha = 0\) case needs to be handled separately, and it corresponds to the geometrically-spaced grid.

\begin{figure}[ht]
\centering
\includegraphics[width=\defaultwidth]{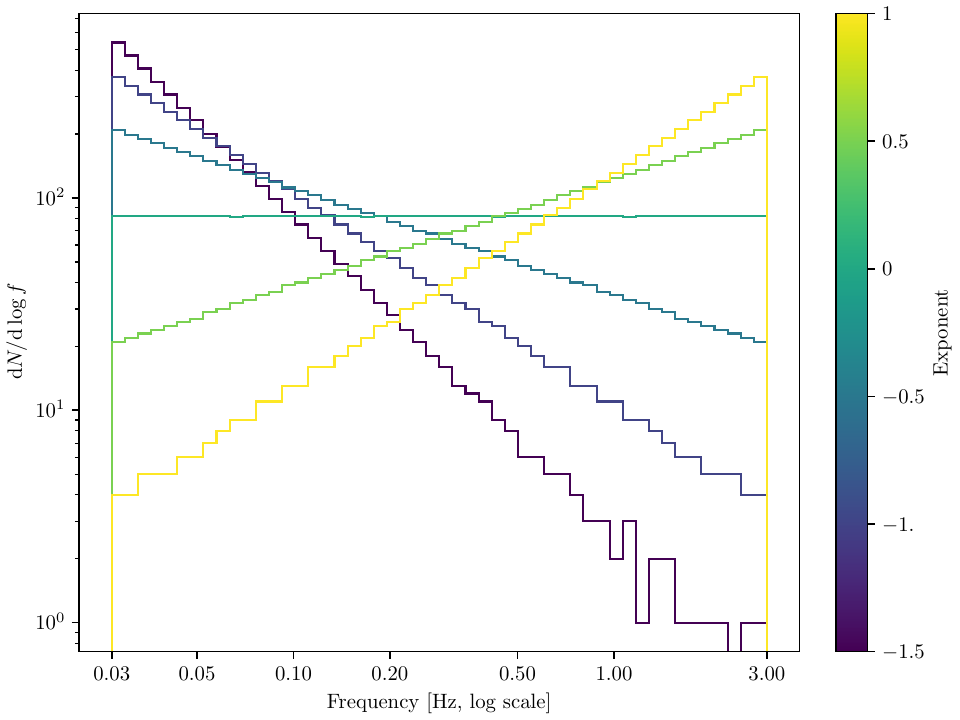}
\caption{Histograms of the polynomial frequency grids with exponents ranging from -1.5 to 1. In all cases, we distribute 4096 grid points according to the given polynomial index, and show their number density as a function of frequency, on a logarithmic axis.}
\label{fig:frequency-grid-histograms}
\end{figure}

The term ``polynomial'' is used due to the shape of the logarithmic histograms of the grids, shown in Figure \ref{fig:frequency-grid-histograms}.

\begin{figure}[ht]
\centering
\includegraphics[width=\defaultwidth]{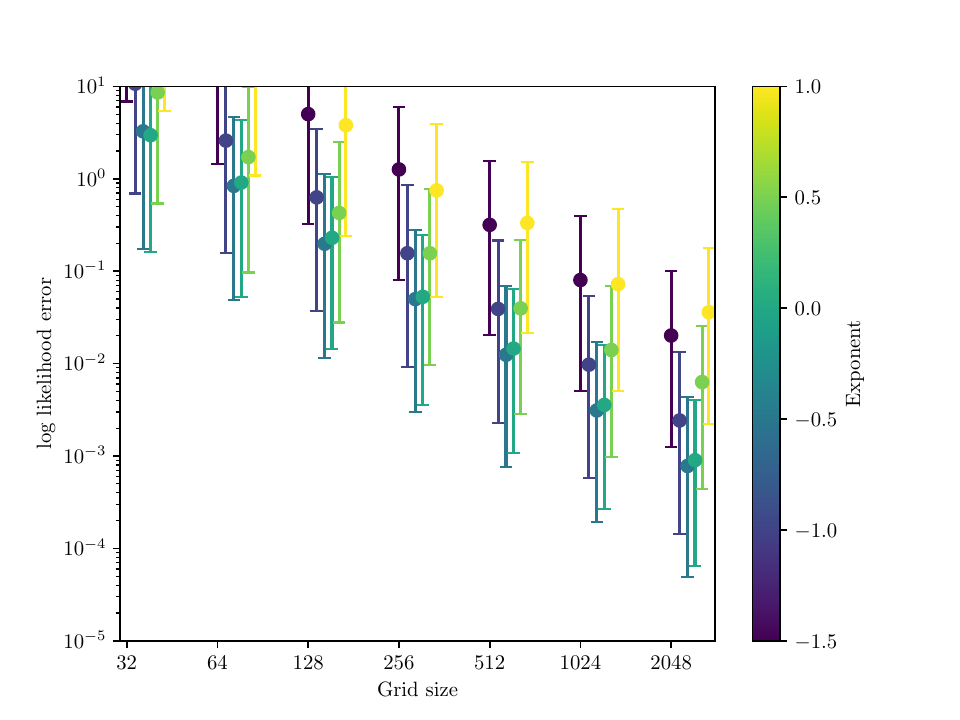}
\caption{Likelihood convergence as a function of grid size. Results are computed for the full GW250114 injection. For each grid size, we estimate the error on the log likelihood by comparing its value to that computed with the next highest resolution. For each scenario, we show the median and the (5, 95)\% quantiles, as computed on the set of posterior samples.
We offset the points for visual clarity, but within each cluster they correspond to the same grid size shown as a label on the horizontal axis.}
\label{fig:likelihood-convergence-by-grid-size}
\end{figure}

As Figure \ref{fig:likelihood-convergence-by-grid-size} shows, likelihood convergence is fastest for values of \(\alpha\) between \(-1\) and \(0\), corresponding to grids with slightly higher density at low frequencies than the geometrically spaced one.

For our main GW250114 injection, we use a 2048-point grid with an exponent of \(-0.5\); for the injections in section \ref{sec:localization-function-area} we use a grid with \(200 n_{\text{months}}\) points and an exponent of \(-0.5\).

\section{Einstein Telescope injections}\label{app:et-injections}

In order to provide a point of comparison, we simulate a detection of GW250114 with the Einstein Telescope. 
As its geometric configuration is not yet certain, 
we consider two different options: a detector shaped like an equilateral triangle with an side length of 10km ("ET-\(\Delta\)"), and a pair of L-shaped detectors, misaligned by 45\(^{\circ}\), at a relative distance of a little over 1000km.
More details on the configurations and further comparisons can be found in reference \cite{santoliquidoComparingNextgenerationDetector2025}, where these are labelled ``ET-\(\Delta\)'' and ``ET-2L-MisA'' respectively.

We work in the frequency domain, in the range \([6, 2048] \ \text{Hz}\), which corresponds to a duration of less than 32 seconds.

We use the same functional form for the priors as the \ac{lgwa} analyses, but with different widths in some parameters.
This has no effect on the posterior distributions, as the differences are in regions far from any prior support. 
In particular: our priors are wider in chirp mass (between \(30\) and \(32\ M_\odot\)) but narrower in mass ratio (between \(0.95\) and \(0.99\)). 
Our distance priors are uniform in the source frame in the range \([400, 450]\text{Mpc}\). 
We use merger time at the geocenter as a parameter, with a uniform prior of width 0.2 s around the injected value.
The priors on all other parameters are identical to the ones discussed in section \ref{sec:methods}.

We employ nested sampling with \texttt{bilby}, the \texttt{nessai} sampler and 5000 live points. Unlike the \ac{lgwa} analyses, we found its default settings to perform sufficiently well here.

The total (component) optimal \ac{snr} for the two configurations is 682 (494, 219, 415) for ET-$\Delta$ and 835 (588, 593) for ET-2L respectively.
The difference is driven by the configurations' arm lengths (10 and 15 km respectively), opening angles (60 and 90 degrees respectively) and the injected source's orientation with respect to the antenna patterns of either configurations, which are different both due to their geometry and locations.

Due to its higher \ac{snr}, the ET-2L configuration exhibits slightly tighter constraints for this event; however, the largest difference between the two configurations is given by the multimodalities. 
The ET-2L posterior distribution is unimodal in all parameters and nearly Gaussian.
The ET-\(\Delta\) one, on the other hand, is bimodal: specifically, about 5\% of the posterior mass lies in a secondary mode centered around the inclination angle \(\theta_{JN} = \pi - \theta_{JN}^{\text{inj}}\), polarization \(\psi = \pi - \psi^{\text{inj}}\), geocentric time nearly 8 ms before the injected value, and sky position reflected across the plane of the local horizon at the Einstein Telescope's position (see Figure \ref{fig:sky-loc-constraint}).

\begin{figure}[ht]
\centering
\includegraphics[width=\defaultwidth]{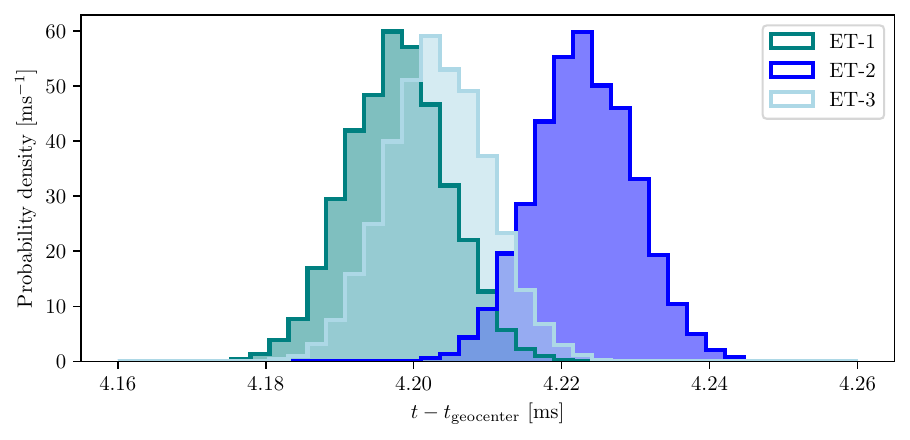}
\caption{Merger times measured at the three vertices of a triangular Einstein Telescope.}
\label{fig:et-triangulation}
\end{figure}

For signals observed by a point-like, stationary triangular Einstein Telescope, an eight-fold multimodality in sky position is observed \cite{santoliquidoFastAccurateParameter2025} (see also \cite{marsatExploringBayesianParameter2021} for a discussion of this in the context of \ac{lisa}).
However, in our model of the Einstein Telescope, the three components are located at three different positions, in an equilateral triangle with a side length of 10 km. The light travel time between them is on the order of tens of microseconds: at this \ac{snr}, this is enough to triangulate the signal, as we show in Figure \ref{fig:et-triangulation}.

The time delays between components are sufficient to exclude six out of the eight modes in the sky, but two remain: specifically, those corresponding to the true source position and its reflection across the local horizon (see also equation 63 in \cite{marsatExploringBayesianParameter2021}).
The time delays corresponding to these two locations in the sky are exactly identical, as they only depend on the projection of the gravitational source vector onto the local horizon plane.
Therefore, the only way they can be distinguished is by measuring the inclination angle well enough to exclude the scenario in which the source is oriented ``upside down'', \emph{i.e.}\ with \(\theta_{JN} = \pi - \theta_{JN}^{\text{inj}}\). 
This discrimination can be achieved thanks to higher mode content in the signal, which we did include for our \ac{et} injection, but it was not sufficient to confidently exclude the second mode.




\section{Full angular resolution calculations} \label{app:angular-resolution-derivation}

In this section, we derive the sky-localization error given in Equation~\ref{eq:wen-chen-area}. This result was first presented in similar form in Ref.~\cite{wenGeometricalExpressionAngular2010}; we rederive it and expand upon it, giving more details regarding its application in a realistic scenario. 

Before starting with the proper discussion it is necessary to give some definitions. 

\begin{description}
    \item[Definition 1:] Inner product of two vectors representing time series recorded with a network of $N$ detectors with noise power spectral density $S_k(f)$ is defined by:
        \begin{equation}
            \langle \vec x|\vec y\rangle:=4\sum\limits_{k=1}^N\int\limits_0^\infty\drm f\frac{\Re[\tilde x_k(f)\tilde y_k(f)^*]}{S_k(f)}
        \label{eq:innProd}
        \end{equation}
    \item[Definition 2:] Schur's complement of a 2 by 2 matrix $ \mathcal{F}=\left(
    \begin{matrix}
        \mathcal{F}_{A} & \mathcal{F}_{B} \\
        \mathcal{F}_{C} & \mathcal{F}_{D}
    \end{matrix}
    \right)$:
        \begin{equation}
            \mathcal{F/F_{D}} :=\mathcal{F}_{A} - \mathcal{F}_{B}\mathcal{F}_{D}^{-1}\mathcal{F}_{C}
        \end{equation}
\end{description}

Now, given a set of parameters ($\Omega, \vec\lambda$), where $\Omega = (\theta, \phi)$ are the sky-localization parameters and $\vec\lambda$ are other $p$ additional parameters, assume to have a generic Fisher matrix:
\begin{equation}
    \mathcal{F}=\left(
    \begin{matrix}
        \mathcal{F}_{\Omega\Omega} & \mathcal{F}_{\Omega\vec\lambda} \\
        \mathcal{F}_{\vec\lambda\Omega} & \mathcal{F}_{\vec\lambda\vec\lambda}
    \end{matrix}
    \right)
\end{equation}
where $\mathcal{F}_{\Omega\Omega}$ is a 2$\times$2 matrix, $\mathcal{F}_{\vec\lambda\vec\lambda}$ is a $p\times p$ matrix, and $\mathcal{F}_{\Omega\vec\lambda}$ and $\mathcal{F}_{\vec\lambda\Omega}$ are appropriately sized.
The components of the matrix are defined as:
\begin{equation}
    \mathcal{F}_{ij}=\langle\partial_i\vec h|\partial_j\vec h\rangle \hspace{10pt} \mathrm{where} \hspace{5pt} i,j = \{\theta, \phi, \vec\lambda\}
    \label{eq:fisherComp}
\end{equation}
Then, its Schur's complement would take the following form:
\begin{equation}
    \mathcal{F/F_{\vec\lambda\vec\lambda}} :=\mathcal{F}_{\Omega\Omega} - \mathcal{F}_{\Omega\vec\lambda}\mathcal{F}_{\vec\lambda\vec\lambda}^{-1}\mathcal{F}_{\vec\lambda\Omega}
    \label{eq:schurCompl}
\end{equation}

The aim of this section is to evaluate the parameter-estimation errors for a subset of parameters, by exploiting the known connection between the Fisher matrix and the covariance matrix described in Equation~\ref{eq:covFisherRelation}:
\begin{equation}
    V_{ij} = (\mathcal{F}^{-1})_{ij}
    \label{eq:covFisherRelation}
\end{equation}
However, to get the covariance matrix of a subset of parameters it is necessary to estimate the Fisher matrix just for the subset of parameters. To do so, we use the properties of Schur's complement, which would enable us to isolate the parameters of interest from the non-relevant ones, as described in Equation~\ref{eq:schurCompl}.

Since we are interested in evaluating the angular resolution, using  Equation~\ref{eq:schurCompl}, it is possible to isolate the Fisher matrix for the subset of parameters $\Omega$:
\begin{align}
    \nonumber \mathcal{F'} &= \mathcal{F}_{\Omega\Omega} - \mathcal{F}_{\Omega\vec\lambda}\mathcal{F}_{\vec\lambda\vec\lambda}^{-1}\mathcal{F}_{\vec\lambda\Omega} = \\
    \nonumber &= \langle \partial_\Omega \vec h| \partial_\Omega \vec h \rangle - \langle \partial_\Omega \vec h| \partial_{\vec \lambda} \vec h \rangle \langle \partial_{\vec \lambda} \vec h| \partial_{\vec \lambda} \vec h \rangle^{-1} \langle \partial_{\vec \lambda} \vec h| \partial_\Omega \vec h \rangle = \\
    &= \langle \partial_\Omega \vec h| \mathbb{I} - \mathbf{P} | \partial_\Omega \vec h \rangle
    \label{eq:schurproject}
\end{align}
where $\mathbb{I}$ is the identity matrix, and $\mathbf{P}$ is defined as $\mathbf{P} := |\partial_{\vec \lambda} \vec h \rangle \langle \partial_{\vec \lambda} \vec h| \partial_{\vec \lambda} \vec h \rangle^{-1} \langle \partial_{\vec \lambda} \vec h| $. The sky-localization error can now be written as
\begin{equation}
    \Delta\Omega=\frac{2\pi}{\sqrt{\det(\mathcal F^\prime)}}.
    \label{eq:subsetFisher}
\end{equation}

To better understand the physical meaning of Equation~\ref{eq:subsetFisher}, we can consider the following simple example of a detector network observing a signal from a plane wave
\begin{equation}
    h_I = A_I(\omega)\exp[\irm(\omega \tau-\vec k(\theta,\phi)\cdot \vec r_I)],
    \label{eq:simpl_signal}
\end{equation}
where $\tau$ is a reference time that determines the phase of the wave at the origin of the coordinate system, $\vec r_I$ are the positions of the detectors, $A_I(\omega)$ the amplitudes of the signals observed at the detectors, and $\theta,\,\phi$ are the angles of the sky-location of the source. 
In the following, we will only include $\tau$ under the model parameters $\vec \lambda$. We then obtain
\begin{equation}
\begin{split}
    \partial_\theta h_I &= [- i \partial_\theta (\vec k \cdot \vec r_I)] h_I  \\
    \partial_\phi h_I &= [- i \partial_\phi (\vec k \cdot \vec r_I)] h_I \\
    \partial_\tau h_I &= [i \omega] h_I
\end{split}
\end{equation}

\begin{equation}
    \mathcal{F}_{\Omega\Omega}^I = \langle \partial_\Omega \vec h_I| \partial_\Omega \hat h_I \rangle = \langle \partial_\Omega (\vec k \cdot \vec r_a) \hat h_a| \partial_\Omega (\vec k \cdot \vec r_a) \hat h_a \rangle
\end{equation}

\begin{equation}
    \nonumber \mathcal{F}_{\Omega \tau}^I = \langle \partial_\Omega h_I| \partial_\tau h_I \rangle  = - \langle \partial_\Omega (\vec k \cdot \vec r_I) h_I| \omega h_I \rangle
\end{equation}

\begin{equation}
    \nonumber \mathcal{F}_{\tau\Omega}^I = \langle \partial_\tau h_I|\partial_\Omega h_I \rangle  = - \langle \omega h_I|\partial_\Omega (\vec k \cdot \vec r_I) h_I \rangle
\end{equation}
\begin{align}
    \nonumber \mathcal{F}_{\tau\tau}^I = \langle \omega h_I|\omega h_I \rangle \quad .
\end{align}
Thus, Equation~\ref{eq:schurproject} becomes:

\begin{equation}\label{eq:Fprime}
    \begin{split}
        \mathcal{F'} & = \sum\limits_{I=1}^N\langle \partial_\Omega (\vec k \cdot \vec r_I) h_I| \partial_\Omega (\vec k \cdot \vec r_I) h_I\rangle- \frac{\left(\sum\limits_{I=1}^N\langle \partial_\Omega (\vec k \cdot \vec r_I)h_I |\omega h_I \rangle\right)\otimes\left(\sum\limits_{I=1}^N\langle \partial_\Omega (\vec k \cdot \vec r_I)h_I |\omega h_I \rangle\right)}{\langle \omega \vec h | \omega \vec h \rangle}
    \end{split} \, .
\end{equation}

We now introduce some notation to simplify the expressions. Writing the wave vector as $\vec{k}=\omega/c\hat n(\theta,\phi)$, where $\hat n$ is the unit vector pointing along the direction of propagation of the wave, we define
\begin{equation}
    \frac{\partial\hat n}{\partial\theta}\equiv \hat n_\theta,\quad \frac{\partial\hat n}{\partial\phi}\equiv \sin(\theta)\hat n_\phi, \quad \frac{\partial\hat n}{\partial\Omega}= (\hat n_\theta,\hat n_\phi)^\top\equiv n_\Omega,
    \label{eq:partial_k}
\end{equation}
where $\hat n_\theta,\,\hat n_\phi$ are unit column vectors in 3D space.

\subsection{Variational approach for long observations}

An alternative approach for obtaining the desired submatrix is the variational one \cite{wenGeometricalExpressionAngular2010}. 
With it, the Fisher matrix pertaining to $\Omega$ could be rewritten as 
\begin{equation}
    \min_{\delta \tau} \langle \delta \vec{h}|\delta \vec{h}\rangle  = \frac{1}{2}\mathcal{F}_{ij}\delta \Omega_i\Omega_j + \mathcal{O}(|\delta\Omega|^3)
    \label{eq:FisherExpand}
\end{equation}

where \begin{equation}
    \delta \vec h \equiv \vec h\left(\Omega+\delta\Omega,\tau+\delta\tau\right)-\vec h\left(\Omega,\tau\right) \,.
\end{equation} 

When considering the case of \textit{long signal} with known waveform, the signal stays in band longer and it is the area of the detector network that determines the angular resolution.

Starting from the signal in Equation~\ref{eq:simpl_signal}, we rewrite the signal variation $\delta\vec h$ as
\begin{equation}
    \delta h_I\approx\left[-\delta \tau_0-\sum_i\frac{r^i_I}{c}\delta n_I\right]\dot{h}_I(t) \, .
\end{equation}

Here we assume that the antenna pattern changes at a much lower frequency than the signal and that the speed of the detector is smaller than the speed of light. The change in antenna beam patterns induced by $\delta n$ is also ignored. 

We now evaluate the product 
\begin{equation}
    \langle\delta h_I|\delta h_I\rangle =\int_0^T\dot{\xi}_I(t)\left[-\delta \tau_0-\sum_i\frac{r^i_I}{c}\delta n_I\right]^2\mathrm{d}t
\end{equation}
 where
\begin{subequations}
\begin{align}
    \label{eq:xi_def}
    \xi_I&=4\int_{0}^\infty \omega^2 \frac{|h_I|^2 }{ S_I(\omega)} \frac{\mathrm{d}\omega}{2\pi}=\langle\omega h_I|\omega h_I\rangle \,,\\
    \label{eq:der_xi_def}
    \dot{\xi}_I&=2\int \dot{h}_I(t-\tau/2)w_I(\tau)\dot{h}_I(t+\tau/2)\mathrm{d}\tau\,,
    \end{align}
\end{subequations}
and $w_I$ is the Fourier transform of $1/S_I(\Omega)$. 

Following the variational approach described in Equation~\ref{eq:FisherExpand}, the Fisher matrix obtained by minimizing over $\delta \tau_0$ has the form:
\begin{equation}
    \mathcal{F}_{ij}=\frac{1}{c^2}\left[\overline{r_ir_j}-\bar{r}_i\bar{r}_j\sum_I\xi_I\right]
\end{equation}

where the average of the $j$-th coordinate of detectors in the network, throughout the detection time, and correlations between the $i$-th and $j$-th coordinates are evaluated as
\begin{subequations}\label{eq:AvCorrCoordinates}
    \begin{align}
        \overline{r_i} &\equiv \frac{\sum_J \int_0^T r_i^J(t) \dot{\xi}_J(t) \, dt}{\sum_J \xi_J(T)}\\
        \overline{r_ir_j} &\equiv \frac{\sum_J \int_0^T r_i^J(t) r_j^J(t) \dot{\xi}_J(t) \, dt}{\sum_J \xi_J(T)}\, .
    \end{align}
\end{subequations}

The quantity $\dot{\xi}_I$ can effectively be seen as a weighting function for the average and correlations evaluated above. 

In this work we operate in the frequency domain, hence we cannot compute \(\dot{\xi}(t)\) directly.
For the purpose of the computations in Sec.~\ref{sec:localization-function-area} and Fig.~\ref{fig:areas-scaling}, we approximate it as follows (see also Eq.~\ref{eq:weight_xidot}):
\begin{equation}
\dot{\xi}(t) \approx w(t) = \left|\frac{\mathrm{d} f}{\mathrm{d}t}\right| \frac{\left|h_x(f(t))\right|^2  + \left|h_y(f(t))\right|^2}{S_n(f(t))} \,,
\end{equation}
where \(f(t)\) is the monotonic time-frequency mapping from the Stationary Phase Approximation, while \(h_x\) and \(h_y\) are the strains from the two \ac{lgwa} channels.

The determinant of the inverse matrix yields the following angular resolution
\begin{equation}\label{eq:AngResWenChen_49}
    \Delta\Omega =\frac{2\pi c^2}{\sum_I\xi_I(T) \sqrt{\mathrm{Var}(r_\theta)\mathrm{Var}(r_\phi)-\mathrm{Cov}(r_\theta,r_\phi)^2}} \,. 
\end{equation}

Before providing any other assumptions, we can look at the denominator in more detail, recovering further simplifications.

The first term appearing in the denominator of Equation~\ref{eq:AngResWenChen_49} is $\sum_I \xi_{I}$, where $\xi_I$ is the same function defined in Equation~\ref{eq:xi_def}. 

The sum can be rewritten as
\begin{equation}
  \sum_I \xi_{I}=\sum_I 4 \int_{0}^\infty \omega^2 \frac{|h_I|^2}{S_I(\omega)}   \frac{\mathrm{d}\omega}{2\pi}=\rho_T ^2 \langle\omega^2\rangle \, ,
\end{equation} 
where \begin{equation}
    \rho_T^2=\sum_J 4 \int_{0}^\infty \frac{|h_J|^2}{S_J(\omega)}   \frac{\mathrm{d}\omega}{2\pi}
\end{equation}
and 
\begin{equation}
    \langle\omega^2\rangle=\left(\sum_J\int_{-\infty}^\infty \omega^2 \frac{|h_J|^2 }{ S_J(\omega)} \mathrm{d}\omega\right)\left(\sum_J\int_{-\infty}^\infty \frac{|h_J|^2 }{ S_J(\omega)}\mathrm{d}\omega\right)^{-1} \,.
\end{equation}

\subsection{Area dependence}\label{subsec:area}
The other term in the denominator of Equation~\ref{eq:AngResWenChen_49} is related to the area mapped out by the trajectory of the network. 
In the following, we are going to define the \textit{statistical area} as in Equation~\ref{eq:statistica_area}, which we rewrite here for convenience

\begin{equation}
    A_s(T)= 2\pi \sqrt{\mathrm{Var}(r_\theta)\mathrm{Var}(r_\phi)-\mathrm{Cov}^2(r_\theta,r_\phi)}\, ,
\end{equation}

where Var and Cov are variance and co-variance with expectation values calculated according to Equations~\ref{eq:AvCorrCoordinates}.

This area is related, but not equivalent, to the \textit{geometrical} area covered by the detector. 
In general, it is possible to write their relation as follows
\begin{equation} \label{eq:statistical-geometrical-ratio}
    A_s= \alpha(T) A_g\,,
\end{equation}
where $\alpha(T)$ is a function of the period of observation and depends on the trajectory of the orbit.
In the case of equally-weighted orbits we generally observe \(\alpha (T) \approx 1\); on the other hand, if most of the \ac{snr} is accumulated in a short fraction of the orbit we will have \(\alpha (T) \ll 1\). 

\begin{figure}[ht]
\centering
\includegraphics[width=\columnwidth]{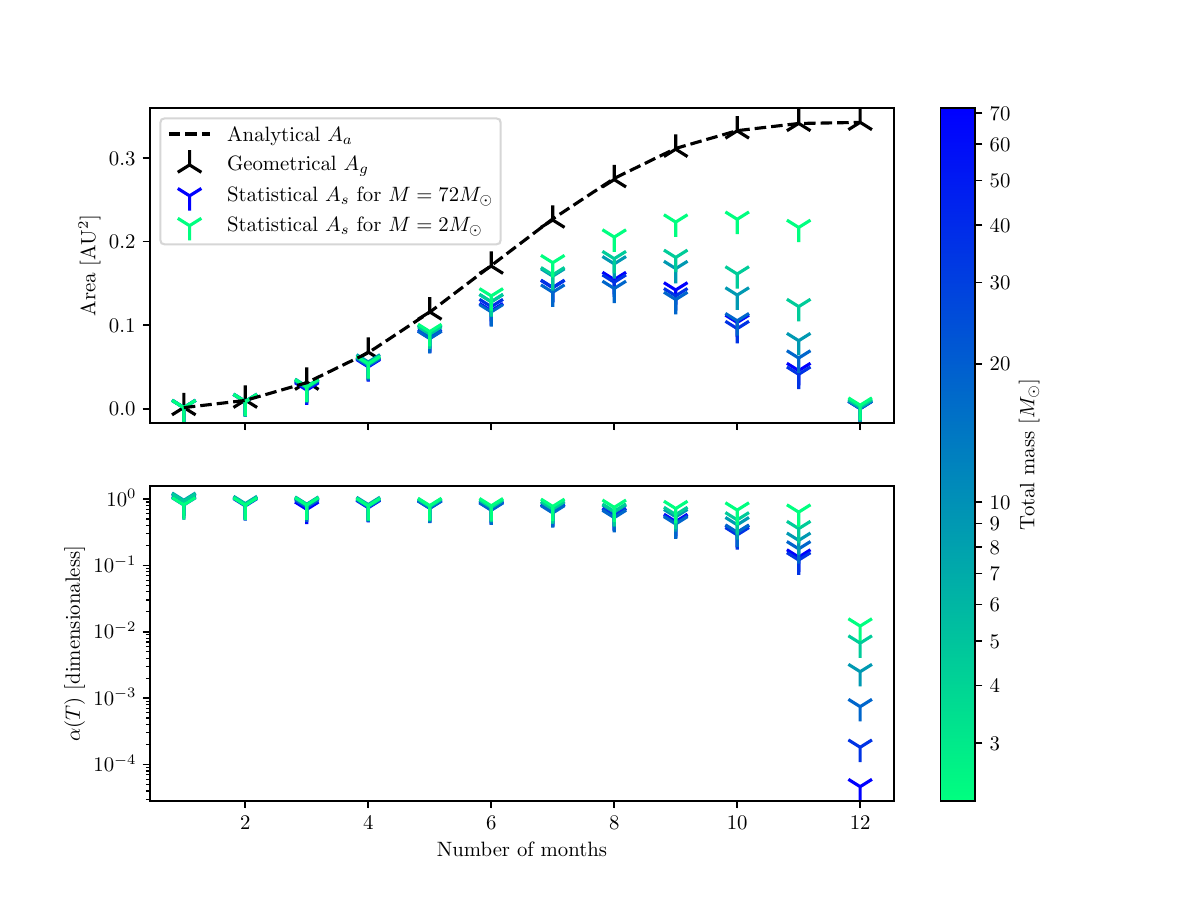}
\caption{Top panel: scaling of the notions of orbital area described in this section, as a function of the number of observation months. As in Sec.~\ref{subsec:area}, we compute the statistical area by keeping the lower frequency boundary fixed and varying the upper one to obtain an observation with the desired number of months. For reference, we also show the analytical approximation \(A_a\) (Eq.~\ref{eq:analytical-area}).
We illustrate how the statistical area \(A_s\) depends on the \ac{snr} accumulation of the signal by repeating the analysis varying (only) the total mass of the system.
Bottom panel: same as the top, but instead of absolute areas we show the ratio \(\alpha(T)\) (Eq.~\ref{eq:statistical-geometrical-ratio}) on a logarithmic scale.}
\label{fig:areas-scaling}
\end{figure}

In Fig.~\ref{fig:areas-scaling} we show the scaling of \(A_s\) and \(A_g\) for the injections considered in Sec.~\ref{subsec:area}. 
The geometric area \(A_g\) can be accurately modelled by a simple analytical formula:
\begin{equation} \label{eq:analytical-area}
A_a = 0.5 (\omega_* t - \sin(\omega_* t)) |\hat{k} \cdot \hat{n}_{\text{ecl}}|  \,,
\end{equation}
where \(\omega_*\) is the orbital frequency of the Earth around the Sun, \(\hat{k}\) is the \ac{gw}'s propagation direction and \(\hat{n}_{\text{ecl}}\) is the mean normal vector to the ecliptic plane. For GW250114, \(\hat{k} \cdot \hat{n}_{\text{ecl}} \approx -0.11\).
We note that for a source with a uniformly random position in the sky we expect \(|\hat{k} \cdot \hat{n}_{\text{ecl}}| \sim \mathcal{U}(0, 1)\): GW250114 was on the lower end of this distribution, hence we expect comparable sources to be better localized on average.

\subsection{General equation}

We can now write Equation~\ref{eq:AngResWenChen_49} as 
\begin{equation}
\label{eq:simplified_49}
    \Delta\Omega=\frac{ (2\pi)^2 c^2}{\rho_T^2\langle\omega^2\rangle A_s(T)}\,,
\end{equation}
from which the monochromatic case in Eq.~\ref{eq:wen-chen-area} directly follows.

We stress here that this result models the posterior distribution on the sky position vector, marginalized \emph{only} over the arrival time parameter. 
While the intrinsic parameters of the source are expected to be broadly uncorrelated with the extrinsic ones \cite{singerRapidBayesianPosition2016}, this does not hold for the other extrinsic parameters, such as phase, polarization angle and distance.

Indeed, in Sec.~\ref{sec:localization-function-area} we observe that the prediction fails to capture the precise magnitude of the sky localization area.
The comparison we make is not exactly one-to-one: the simplest case we explore with full parameter estimation considers four parameters, including phase beyond the ones considered here.
Failing to marginalize over phase leads to pathological posterior distributions, with severe multimodalities in the timing parameter.

Nevertheless, the prediction \(\Delta \Omega \propto 1 / A_s\) does hold, making this Fisher analysis a useful tool for understanding the localization capabilities of orbiting detectors.

\subsubsection{On Reference Frame Independence} 

In this subsection, we stress the importance of getting a formula for the sky localization error which is independent of the reference frame.

From Equation~\ref{eq:simplified_49}, this independence is evident, as the quantity $A_s$ is computed through the variance and co-variance of the detector motion vector.

On the other hand, from Equation~\ref{eq:Fprime} there seems to be an ``apparent'' dependence of the Fisher matrix on the coordinate system.

In the case that the detectors do not (significantly) change their position while a signal is observed, we have
\begin{equation}
    \begin{split}
        \mathcal{F'} & = \sum\limits_{I=1}^N(n_\Omega \cdot \vec r_I)\otimes(n_\Omega \cdot \vec r_I)/c^2\langle \omega h_I| \omega h_I\rangle\\
        &\qquad\qquad- \frac{\left(\sum\limits_{I=1}^N(n_\Omega \cdot \vec r_I)/c\langle \omega h_I| \omega h_I\rangle\right)\otimes\left(\sum\limits_{I=1}^N(n_\Omega \cdot \vec r_I)/c\langle \omega h_I| \omega h_I\rangle\right)}{\langle \omega \vec h | \omega \vec h \rangle}
    \end{split}
    \label{eq:stationary}
\end{equation}
For clarity, we make another simplification by assuming that the observations of the signal are all done with the same sensitivity and the signal has the same amplitude $h$ in each detector. Equation~\ref{eq:stationary} then simplifies to
\begin{equation}
    \begin{split}
        \mathcal{F'} & = \langle \omega h| \omega h\rangle\frac{1}{c^2}\left[\sum\limits_{I=1}^N(n_\Omega \cdot \vec r_I)\otimes (n_\Omega \cdot \vec r_I)-\frac{1}{N}\sum\limits_{I,J=1}^N(n_\Omega \cdot \vec r_I)\otimes(n_\Omega \cdot \vec r_J)\right]\\
        & = \langle \omega h| \omega h\rangle\frac{1}{c^2}\sum\limits_{I=1}^N(n_\Omega \cdot \vec r_I)\otimes\left[(n_\Omega \cdot \vec r_I)-\frac{1}{N}\sum\limits_{J=1}^N(n_\Omega \cdot \vec r_J)\right]\\
        & = \langle \omega h| \omega h\rangle\frac{1}{c^2}\frac{1}{N}\sum\limits_{I,J=1}^N(n_\Omega \cdot \vec r_I)\otimes\left[n_\Omega \cdot (\vec r_I-\vec r_J)\right]\\
        & = \langle \omega h| \omega h\rangle\frac{1}{c^2}\frac{1}{2N}\sum\limits_{I,J=1}^N\left[n_\Omega \cdot (\vec r_I-\vec r_J)\right]\otimes\left[n_\Omega \cdot (\vec r_I-\vec r_J)\right].
    \end{split}
\end{equation}
The determinant of the $2\times 2$ matrix can be calculated exploiting the Cauchy-Binet formula together with the vector identity
\begin{equation}
    (\vec a\cdot\vec c\,)(\vec b\cdot\vec d\,)-(\vec a\cdot\vec d\,)(\vec b\cdot\vec c\,)=(\vec a\times\vec b\,)\cdot(\vec c\times\vec d\,),
    \label{eq:vec_id}
\end{equation}
The result is
\begin{equation}
    \begin{split}
        \det(\mathcal{F'}) = \langle \omega h| \omega h\rangle^2\frac{1}{8N^2c^4}\sum\limits_{k,l,n,m=1}^N\left[\hat n \cdot \left[(\vec r_l-\vec r_m)\times(\vec r_k-\vec r_l)\right]\right]^2,
    \end{split}
\end{equation}
where we used $\hat n=\hat n_\theta\times\hat n_\phi$, which is the unit vector pointing along the direction of propagation of the GW. 
The coordinate independence of this result is now manifest, since it is written only in terms of differences of position vectors.
The minimum number $N$ of detectors that leads to a non-vanishing result is 3 here: while two detectors can localize a source in the sky, this Fisher result is only considering timing measurements, not antenna patterns. 
With these, the posterior distribution is fully degenerate in the direction orthogonal to the inter-detector vector.

\subsection{Detectors on circular orbits}
We now proceed with the evaluation of the angular resolution for a simple case study: detectors on circular orbits, with radius $R_*$ and orbital frequency $\omega_*$.

Assuming that $\dot{\xi}_J=\mathrm{const}$ for each detector, the average and correlation in Equations~\ref{eq:AvCorrCoordinates} can be rewritten as
\begin{subequations} 
\begin{align}
    \overline{r_k} &= \frac{\sum_J \int_0^T r_k^J(t) \, dt}{N T}\\
    \overline{r_jr_k} &= \frac{\sum_J \int_0^T r_j^J(t) r_k^J(t)  \, dt}{N T}\, ,
\end{align} 
\end{subequations}

where $N$ is the total number of detectors and $T$ the total observation time.

In the situation where the network trajectory has mapped out a size much larger than the size of the network and $\omega_*T\gg1$, we can write the detector position vector $\mathbf{r}(t)$ and $\mathbf{n}$ as:
\begin{subequations} 
\begin{align}
    \mathbf{r}(t) &= (R_* \cos \omega_* t, \, R_* \sin \omega_* t, \, 0) \\
    \mathbf{n} &= (\cos \iota, \, 0, \, \sin \iota),
\end{align}
\end{subequations}
where we defined the inclination angle $\iota$ between the detector and the wave propagation direction.
To compute the variance and covariance terms in Equation~\ref{eq:statistica_area}, we first need the derivatives of $\hat{n}$:
\begin{equation}
    \partial_\theta \mathbf{n} = (0, 1, 0), \quad \partial_\phi \mathbf{n} = (-\sin \iota, 0, \cos \iota),
\end{equation}
implying simply:
\begin{subequations} 
\begin{align}
    r_\theta \equiv \partial_\theta \mathbf{n} \cdot \mathbf{r}(t) &= R_* \sin(\omega_* t) \\
    r_\phi \equiv \partial_\phi \mathbf{n} \cdot \mathbf{r}(t) &= -R_* \sin \iota \cos(\omega_* t).
\end{align}
\end{subequations}

Now we use our long signal assumption, $\omega_*T\gg1$, which allows us to compute the variance and covariance terms:
\begin{subequations} 
\begin{align}
    \text{Var}(r_\theta) &= \langle r_\theta^2 \rangle = \frac{1}{2} R_*^2 \\
    \text{Var}(r_\phi) &= \langle r_\phi^2 \rangle = \frac{1}{2} R_*^2 \sin^2 \iota \\
    \text{Cov}(r_\theta, r_\phi) &= \langle r_\theta r_\phi \rangle = 0.
\end{align}
\end{subequations}

Finally, we plug in Equation~\ref{eq:statistica_area} all the terms, getting:
\begin{equation}
    A_s=2\pi\sqrt{\mathrm{Var}(r_\theta)\mathrm{Var}(r_\phi)-\mathrm{Cov}(r_\theta,r_\phi)^2} = 2\pi\sqrt{\left( \frac{1}{2} R_*^2 \right) \left( \frac{1}{2} R_*^2 \sin^2 i_n \right)} = \pi R_*^2 |\sin \iota|,
\end{equation}
which is the expression of the statistical area, which, under these simple assumptions, gives simply the projection of the geometrical area of the orbit.

Finally, for a monochromatic signal $\omega=\omega_0$, the angular resolution becomes:

\begin{equation}
    \Delta\Omega=\frac{ (2\pi)^2 c^2}{\omega_0^2 \rho_T^2 A_s}=\frac{c^2}{f_0^2 \rho_T^2 A_s}=\frac{c^2}{f_0^2 \rho_T^2 A_g}\, .
\end{equation}



\end{document}